\documentclass[preprint,notitlepage,aps,prb,english,nolongbibliography,superscriptaddress,floatfix]{revtex4-2}
\usepackage[colorlinks=true, urlcolor=blue, linkcolor=blue, citecolor=blue, pdftex]{hyperref}
\usepackage{xr}

\usepackage[dvipsnames]{xcolor}
\usepackage{booktabs}
\usepackage[utf8]{inputenc}
\usepackage[normalem]{ulem}
\usepackage{braket}
\usepackage{graphicx}
\usepackage{multirow}
\usepackage{amsmath}
\usepackage{amsfonts} 
\usepackage[capitalise]{cleveref}

\newcommand{\NRI}{\mathrm{NR_I}}
\newcommand{\NRII}{\mathrm{NR_{II}}}

\begin{document}

\title{On Rare Nonresonant Regions and Subdiffusive Transport in an Interacting Disordered Quantum Chain}

\date{Sept 12, 2026}
\author{A.~Scardicchio}
\affiliation{The Abdus Salam ICTP, Strada Costiera 11, 34151 Trieste, Italy}
\affiliation{INFN, Sezione di Trieste, Via Valerio 2, 34127 Trieste, Italy}
\author{S.~L.~Sondhi}
\affiliation{Rudolf Peierls Centre for Theoretical Physics, University of Oxford, Oxford OX1 3PU, U.K.}

\begin{abstract}
We study rare nonresonant regions in a canonical one-dimensional disordered quantum spin chain by directly implementing the iterated Schrieffer–Wolff construction underlying the recent work of De Roeck, Giacomin, Huveneers and Prośniak on subdiffusive transport. For finite systems, we compute the probability that a disorder realization remains nonresonant through successive scales of the flow, while separately monitoring operator proliferation and the spatial localization of dressed local observables.

The survival probability decays at most exponentially with system size over the numerically accessible regime. Resolving this decay scale by scale yields failure rates  associated with successive Schrieffer–Wolff steps. The first rate is obtained analytically  in excellent agreement with numerics, while higher-scale rates decrease rapidly throughout the regime studied. These results support the summability mechanism required for exponentially rare but parametrically long nonresonant regions, which provide the insulating bottlenecks responsible for subdiffusive transport. At the smallest couplings studied, the survival probability under our conditions, exceeds the rigorous lower bound by six to seven orders of magnitude, demonstrating how conservative the constants required by the proof are while confirming that its underlying physical mechanism is quantitatively visible at accessible scales.

More broadly, our results show how direct numerical implementations can provide an independent and physically transparent test of technically demanding constructive proofs, a methodology that may become increasingly useful as machine-assisted proofs become more common.
\end{abstract}

\maketitle 

\section{Introduction}

The problem of many-body localization was posed in P. W. Anderson's seminal 1958 paper \cite{Anderson58}. In that paper itself, he made progress on the question of localization for non-interacting quantum particles, a phenomenon now forever associated with his name. But he had in mind that this progress was a proxy for the real phenomenon, which was the breakdown of transport and thermalization in laboratory interacting systems.

Almost 70 years later, there has been a great deal of interesting thinking on the problem, initially by Anderson and his co-workers, and, beginning with the work of Basko, Aleiner and Altshuler\cite{basko2006metal}, alongside closely related work of Gornyi, Mirlin and Polyakov\cite{gornyi2005interacting},
by a much larger community\cite{oganesyan2007localization,Pal10,bardarson2012unbounded,iyer2013many, de2013ergodicity,serbyn2013local,ros2015integrals,chandran2015constructing,nandkishore2015many,smith2016many,alet2018many,abanin2019colloquium,sierant2025many}. But the step of decisively showing that the phenomenon of many-body localization actually exists has proven difficult  \cite{vsuntajs2020quantum,sels2021dynamical,abanin2021distinguishing}.

Without recapitulating the full history here, we note that, as of today, the central sharp question is whether a generic, strongly disordered, short-range interacting system in one dimension can, for typical disorder realizations, be diagonalized by a quasi-local unitary in the thermodynamic limit. Outside this setting, especially in higher dimensions, there are good arguments that systems can exhibit extremely slow relaxation without genuinely localizing.

A landmark result in the mathematical-physics literature is Imbrie's construction of a quasi-local diagonalization for a one-dimensional disordered spin chain \cite{Imbrie16a,Imbrie16, imbrie2017local}, conditional on a physically motivated assumption limiting level attraction. Nevertheless, the subsequent physics literature has continued to debate the thermodynamic status of MBL, with rigorous multiscale constructions, finite-size numerical studies, and resonance-based arguments approaching the problem in rather different regimes and technical languages. 

Absent consensus on this question, more on which below, one can ask a weaker one: whether there exist rare intervals on which such a quasi-local diagonalization is possible, or, in the jargon of the field, whether there are regions that exhibit well-defined \(\ell\)-bits, or local integrals of motion. 
If the probability of such intervals is bounded below by an exponential in their length, then sufficiently long localized intervals remain common enough in a much longer chain to obstruct normal diffusion of a conserved density.  
This rare-region mechanism was introduced into the physics literature by Agarwal et al.\cite{agarwal2015} and studied in long chains by other techniques by Žnidarič, Scardicchio, and Varma \cite{vznidarivc2016diffusive,panda2019can,schulz2020phenomenology}.

Recently, De Roeck, Giacomin, Huveneers, and Prośniak\cite{DRGHP} (henceforth DRGHP) turned to this weaker problem using the machinery of mathematical physics and proved an exponential lower bound on the probability of such rare localized intervals, and from this the absence of normal heat conduction in a family of one-dimensional disordered spin chains. Their proof constructs, for sufficiently small perturbations about a trivially localized limit and for each nonresonant disorder configuration, a quasi-local diagonalizing unitary by an iterated Schrieffer--Wolff transformation. 

In the present paper, we interrogate the constructive core of their proof by direct numerical computation. For a finite system we follow the successive Schrieffer--Wolff transformations and determine whether a given disorder realization remains nonresonant through successive steps of the flow and thus admits the finite-size analog of a set of local integrals of motion. The computational technology we use represents operators directly as sparse sums of Pauli strings and allows us to address this more bounded question up to system size $L=32$, much bigger than reachable by standard exact diagonalization. The limit $L=32$ comes from the 64-bit encoding of the Pauli strings in our code; a wider encoding would allow, e.g., $L=64$.

An important feature of our calculation is that we can resolve the survival probability by scale. Defining 
$$
S_k(L)
=
\mathbb{P}\!\left(
\text{survive step }k\,\middle|\,\text{survived through step }k-1
\right),
$$

we can write, 

$$
S_k(L)\simeq A_k e^{-a_k L}.
$$

The quantities \(a_k\) are then the rates per spin for resonances that first appear at stage $k$ of the iteration. The total probability of remaining nonresonant is controlled by the product of the \(S_k\)'s, and hence by the sum

$$
a=\sum_{k\ge1}a_k.
$$

If this series is convergent, the probability of a completely nonresonant interval remains exponentially small in its length; if it does not, that probability becomes smaller than exponential and the rare-region mechanism used in the transport argument fails. The ability to define and measure these scale-resolved rates is a major advantage of the numerical construction.

We find that the measured rates decrease rapidly in the perturbative regime. The data presently resolve several successive rates and show a clear hierarchy indicative of convergence over the range of couplings in which the calculation is well controlled. Quantitatively, we find that the rigorous bounds are enormously more conservative than the behavior seen numerically. At the smallest couplings and system sizes where a comparison can be made, the survival probability of our implementation of the SW flow exceeds the lower bound of DRGHP by six to seven orders of magnitude. This is interesting data in its own right: it gives a quantitative measure of how far the constants required for mathematical control lie from the regime in which the same physical mechanism is visible computationally.

At larger couplings, our data suggest that successive resonance rates approach one another, pointing in turn to a finite radius of convergence of the resonance-free Schrieffer–Wolff hierarchy. If this behavior persists to higher scales, it would define a sharp transition in the statistics of rare nonresonant regions. Such a transition need not coincide with the full MBL transition, since the latter permits resonances provided they can be resolved into sparse local blocks.

 We had three reasons for carrying out this exercise. First, as already noted, the computational technology we use allowed us to study this sharply bounded localization question at considerably larger sizes than standard exact diagonalization. Second, the actual magnitude of the non-resonance probability is an interesting quantity to know quantitatively and as noted we find it is vastly bigger than the proof might suggest. Third, by mirroring the constructive part of the proof while using standard physics techniques, we are able to provide an independent path to achieving confidence in its correctness. Further, it allows other physicists to check our check and to push it parametrically further which would boost this confidence further.

There is a broader motivation for this last point, to which we return at the end of the paper. Many-body localization has been unusual in that standard physics methods based on extrapolation from finite systems have proven inconclusive thus requiring technically demanding multiscale mathematics to reach definitive conclusions in the first instance. Historically, confidence in difficult results in physics has often come from arriving at the same conclusion by several substantially independent routes. Direct numerical reconstruction of a constructive proof supplies such a complementary route in this problem. 
Indeed as we enter an era of AI generated proofs, even those checked via AI-assisted formalization, it could become quite important to inspect key parts of the argument via more ``human friendly" techniques and we regard this paper as a useful step in that direction.

The remainder of the paper is organized as follows. We first introduce the model and the De Roeck--Giacomin--Huveneers--Prośniak construction, and explain the relation between their nonresonance conditions and the condition implemented numerically. We then describe the iterated Schrieffer--Wolff flow, its representation in the Pauli-string basis, and the numerical safeguards needed to distinguish a physical resonance from failure caused by computational truncation. We next present the scale-resolved survival probabilities and resonance rates, beginning with the analytically calculable first scale and proceeding to the higher-scale numerical results, and compare them with the bounds available from the rigorous analysis. We then collect the analytical results, the exact first-scale rate $a_1$ and the leading small-$t$ form of $a_2$, together with a toy model of the flow, in Sec.~\ref{sec:analytic}. We conclude with the implications for rare insulating regions and subdiffusive transport, the relation of this weaker problem to full many-body localization, and the broader question of independent numerical and machine-assisted verification of technically demanding mathematical-physics results.

\section{Model and definition of resonance at scale $k$}
\label{sec:model}

Reference \cite{DRGHP}, which is heretofore called DRGHP, establishes two results for the chain
\begin{equation}
  H \;=\; \sum_{x=1}^{L}\theta_x Z_x
  \;+\;\sum_{x=1}^{L-1}\kappa_x Z_x Z_{x+1}
  \;+\!\!\sum_{\substack{I\subset\Lambda_L\\ \text{interval}}}\!\!
  (\gamma/2)^{|I|} W_I ,
  \label{eq:DRGHPmodel}
\end{equation}
with $\theta_x$ i.i.d.\ uniform on $[0,1]$, $\kappa_x$ deterministic and
bounded, $\|W_I\|\le 1$, and $\gamma$ small:
\begin{itemize}
\item \textbf{Theorem 1 (locality of $U$).} With probability at least
  $e^{-\gamma^{c'}L}$, the diagonalizing unitary $U$ is quasi-local: any operator
  supported in an interval $I$ has $UOU^\dagger=\sum_n O_n$ with $O_n$ supported
  in the $n$-neighbourhood of $I$ and $\|O_n\|\le \gamma^{cn}\|O\|$.
\item \textbf{Theorem 2 (absence of normal conduction).} The time-averaged heat
  current obeys $|\langle J^{(L)}\rangle_{\rm ne}|\le C L^{-\frac{c}{5}\log(1/\gamma)}$
  with probability $1-\exp(-L^{1-\gamma^{c'}}/\log L)$, so that
  $\lim_{L\to\infty}\mathbb{E}\,(L\langle J^{(L)}\rangle_{\rm ne})=0$.
\end{itemize}

Theorem 2 follows from Theorem 1, which is what we want to check numerically. To do this, we specify the $W_I$ operators to a single spin flip $X_i$ so our model is:

\begin{equation}
  H = H^{(0)}= E^{(0)} + V^{(0)},\qquad
  E^{(0)} = \sum_i h_i Z_i + \sum_i J_i Z_iZ_{i+1},\qquad
  V^{(0)} = t\sum_i X_i ,
  \label{eq:model}
\end{equation}
with $h_i, J_i$ i.i.d.\ uniform on $[-1,1]$ and $t$ the perturbation strength $\gamma/2$ in DRGHP. The superscript is a book-keeping device for the Schrieffer-Wolff (SW) transformation. 

The SW transformation is used to build the diagonalizing unitary 
\begin{equation}
  U = \lim_{k\to\infty} e^{-A^{(k)}}\cdots e^{-A^{(1)}},
  \qquad
  H^{(k+1)} = e^{A^{(k+1)}} H^{(k)} e^{-A^{(k+1)}},
  \label{eq:flow}
\end{equation}
each step removing the off-diagonal part of the Hamiltonian to progressively
higher order. This is the same object that appears as
a Newton/KAM iteration in \cite{Imbrie16} and in \cite{DRGHP}.

To find the anti-hermitian $A^{(k)}$ at each scale $k$, the Hamiltonian is split
as $H^{(k)}=E^{(k)}+V^{(k)}$, where $E^{(k)}$ collects the terms diagonal in the
$Z$ basis (products of $I$ and $Z$ only) and $V^{(k)}$ the rest. The generator is
defined by the SW condition
\begin{equation}
  \big[A^{(k+1)},E^{(k)}\big] = -V^{(k)}_{\rm per},
  \label{eq:SW}
\end{equation}
where the subscript ${\rm per}$ means the off-diagonal terms at step $k$ are removed. In this way
\begin{equation}
    e^{A^{(k+1)}}H^{(k)}e^{-A^{(k+1)}}=E^{(k)}+V^{(k)}-V^{(k)}_{\rm per}+...=E^{(k+1)}+...\ .
\end{equation}
Each off-diagonal term is an $X$-monomial (the notation of \cite{DRGHP}, their
(4.12))
\begin{equation}
  V_S \;=\; X_S\,f(Z), \qquad X_S=\prod_{x\in S}X_x ,
\end{equation}
labelled by the set $S$ of sites it flips and $\chi_S(i)=0,1$ is the indicator function for the set $S$. Using
$f(Z)X_S = X_S f(\ldots,(-1)^{\chi_S(i)}Z_i,\ldots)$ one has
$[A,E]=-X_S\,a(Z)\,\partial_S E$, where
\begin{equation}
  \partial_S E \;\equiv\; X_S\,E\,X_S - E
\end{equation}
is the change in the diagonal energy under the flip. Hence \eqref{eq:SW} is
solved term by term by
\begin{equation}
  A_S \;=\; \frac{V_S}{\partial_S E} ,
\end{equation}
which is Eq.~(7.5) of \cite{DRGHP}.

Two identities make this cheap to compute in the Pauli basis.

\paragraph{The denominator is a signed sum over anticommuting $Z$-strings.}
For a diagonal term $c_{S'}Z_{S'}$, conjugation by $X_S$ multiplies it by
$(-1)^{|S'\cap S|}\equiv\prod_i(-1)^{\chi_{|S'\cap S|}(i)}$, hence
\begin{equation}
  \partial_S E \;=\; -2\!\!\sum_{S' \,:\, \{Z_{S'},X_S\}=0}\!\! c_{S'} Z_{S'} ,
  \label{eq:denominator}
\end{equation}
because the bracket $(-1)^{|S'\cap S|}-1$ vanishes for even overlap and equals
$-2$ for odd overlap, and odd overlap is exactly anticommutation. The code
therefore builds the denominator by filtering $E$ with the same branchless
commutation test used everywhere else.

\paragraph{Anti-Hermiticity is automatic.}
Since $X_S(\partial_S E)X_S=-\partial_S E$, the function $1/\partial_S E$ is
\emph{odd} under the flip on $S$, so its Walsh expansion is supported entirely
on $Z$-strings that anticommute with $X_S$. Every such string, multiplied into
$X_S$, produces exactly one factor of $\pm i$. Hence all coefficients of
$A$ in the Hermitian Pauli-string basis are purely imaginary and $A^\dagger=-A$
to machine precision, with no symmetrization step. 

We code the Pauli basis in bit-strings directly and implement the transformations of the bit-strings required in the SW transformation directly in the code. This trick has been used previously in numerical libraries \cite{broers2026exclusive, krotz2026paulib} which in any case we do not use (it is straightforward to code it directly in \texttt{C++} with the help of Claude). 

The generator $A_S$ for a given flip-set, then, is obtained not by inverting a matrix but by inverting the diagonal operator $\partial_S E$ pointwise, the transformation between its two representations being carried out by a fast Walsh--Hadamard transform. For an
off-diagonal term $V_S=X_S f(Z)$, the denominator $\partial_S E$ is diagonal in
the $Z$ basis and, supported on the $Z$-strings that
anticommute with $X_S$; these are collected with the same branchless
commutation test used throughout. Let $\{i_1,\dots,i_m\}$ be the union of their
supports. On a background configuration $\sigma$ restricted to these $m$ sites
each such string $Z_{S'}$ has eigenvalue $(-1)^{S'\cdot\sigma}$, so that
$\partial_S E(\sigma)=-2\sum_{S'}c_{S'}(-1)^{S'\cdot\sigma}$. This map from the
$2^m$ coefficients $c_{S'}$ to the $2^m$ values $\partial_S E(\sigma)$ is an
inverse Walsh--Hadamard transform: the $Z$-strings play the role of Fourier
frequencies and the configurations that of positions. A single in-place
$O(m2^m)$ butterfly\footnote{A butterfly is the elementary step of the fast transform: a pair of entries $(a,b)$ is replaced by $(a+b,a-b)$, the name coming from the crossing lines of its data-flow diagram. Since the $2^m\times2^m$ Walsh--Hadamard matrix is the $m$-fold tensor product of $\big(\begin{smallmatrix}1&1\\1&-1\end{smallmatrix}\big)$, the transform is done in $m$ stages; at stage $j$ every entry is combined by a butterfly with the one whose index differs only in bit $j$. This takes $m\,2^{m-1}$ additions and subtractions, overwriting the input (in place).} therefore yields the denominator on every background at
once, in place of the $O(4^m)$ cost of evaluating each configuration separately.

Because the operator is diagonal, its inversion is the reciprocal of a number on
each configuration, $g(\sigma)=1/\partial_S E(\sigma)$. Here that the
resonance criterion acts: if any $|\partial_S E(\sigma)|$ is anomalously small
the rotation angle $|c_S/\partial_S E(\sigma)|$ exceeds unity and the flow is
declared resonant, rather than a near-singular denominator being inverted. The
coefficients of $A_S$ are then recovered by transforming $g(\sigma)$ back to the
$Z$-string basis with a forward Walsh--Hadamard transform; multiplication by
$X_S$ and by the amplitude $c_S$ gives $A_S=X_S\,c_S\,g(Z)$, with the phase
generated by $X_S Z_{S'}$ rendering it anti-Hermitian, as the oddness of
$\partial_S E$ under the flip guarantees. The construction is exact on the $m$
sites the denominator spans; since $E^{(k)}$ spreads as the flow proceeds, $m$
grows with the scale, and it is this growth that the support cutoff $M_{\max}$
bounds, so that the transform never exceeds $2^{M_{\max}}$.

With $A^{(k+1)}$ in hand, \eqref{eq:flow} is evaluated by a truncated
Baker--Campbell--Hausdorff series. Writing $H_1=[A,H]$ and
$H_n=[A,H_{n-1}]/n$, one has $H_n=\mathrm{ad}_A^n H/n!$ and
\begin{equation}
  H^{(k+1)} = H^{(k)} + \sum_{n\ge1} \frac{\mathrm{ad}_A^n H^{(k)}}{n!}.
\end{equation}
The cancellation of $V_{\rm per}$ at first order is automatic. Rather than recompute $[A,E]$, the engine reuses the by-product $C$ of the generator construction: with $C_E(x)=-1$ wherever the denominator was retained and $0$ elsewhere, $C=X_S\,c_v\,C_E$ equals $-V$ on exactly the configurations that were rotated, so that the perturbation is cancelled where it should be and left in place where it was not. As a check on the internal
consistency of the encoding: $C_E$ is \emph{even} under the flip, so $C$ carries
real coefficients and is Hermitian, as $[A,E]$ must be.

Now we need to define some safeguards for implementing the SW flow. 

{\it 1)} The first is a cutoff $\epsilon$. Consider an operator whose representation over the set of Pauli strings $P$ is $O=\sum_{P}c_P P$. Any string whose amplitude is $|c_P|<\epsilon$ is removed from the array (and its weight recorded to be checked at the end). This is true for all operators, be them diagonal (like $E^{(k)}$) or off-diagonal (like $V^{(k)}$). The same threshold is applied pairwise inside the commutator, where a product with $4|c_a|^2|c_b|^2<\epsilon^2$ is never formed. The production runs use $\epsilon=10^{-6}$; the dependence of the results on $\epsilon$ is discussed below.

{\it 2)} The diameter $D(O)$ of an operator $O$ is the size of the largest Pauli string in the operator. While $D\leq L$, we choose $D_{\rm{max}}=10$ uniformly in $L$. In the perturbative region the amplitudes decay exponentially with the size of the string so this is supposed to truncate amplitudes $|c_P|\sim t^{D_{\rm max}}\ll \epsilon$ anyway. For the denominators this translates in a separate cutoff parameter, which we call $M_{\rm{max}}$. Every diagonal term of $E^{(k)}$ anticommuting with $X_S$ contributes to
$\partial_S E^{(k)}$, and after a few scales these include long strings of negligible weight that inflate the support over which the $2^m$ background configurations must be evaluated. We therefore accumulate the anticommuting terms in order of decreasing $|c_{S'}|$ until their joint support would exceed
$M_{\max}=16$ sites, recording the omitted tail. This is the numerical counterpart of \S7.2 of \cite{DRGHP}; since every retained term still anticommutes with $X_S$, the denominator remains odd under the flip and the
generator exactly anti-hermitian.

{\it 3)} Call $n_x$ the number of strings which touch site $x$, then $N_{\rm{LOC}}=\max_{x=1,...,L} n_x$. We stop the flow when $N_{\rm{LOC}}$ becomes larger than a certain number (at least 6000). The stopped flow is declared {\it undecided} since we are not able to decide whether it converges or not. The realizations which are undecided are re-run with a larger $N_{\rm{LOC}}$, along the ladder $N_{\rm{LOC}}=9000\to2\times10^5\to10^6\to5\times10^6$, until they are resolved into the ``converged" or ``diverged" classes. In the bulk of the data about $20\%$ of the undecided samples turn out to diverge. Among the hardest samples, which need $N_{\rm{LOC}}\geq10^6$, the share is much larger, about $80\%$. Ignoring this fact, and assigning {\it all} the undecided realizations to one class or the other (convergent or divergent) allows us to bound the probability from above and below, and hence the coefficients $a_k$ (see later).

{\it 4)} The rotation $e^{A}He^{-A}=H+\sum_{n\geq1}H_n$ is computed iteratively, $H_1=[A,H]$ and $H_n=[A,H_{n-1}]/n$: each order costs one commutator with $A$, and only the last term is kept. After each order the strings with amplitude below $\epsilon$ are dropped, with the same pairwise threshold inside the commutator as in point 1. Each order is smaller than the previous one by about the rotation angle, so the terms die out and the series ends by itself when no string survives. Over about $6\times10^5$ rotations in our data, the median order at which the series ends is $9$ at the first step of the flow, $5$ at the second, $3$ at the third and $2$ afterwards, and $99\%$ of the rotations end by order $19$. The upper order is therefore not an issue, and we fix it at $120$. This cap is reached in $25$ samples; those samples are flagged and treated as undecided.

\vspace{1cm}

With all these safeguards in place, we run the flow and decide if it converges or diverges. A {\it convergent} flow has $||V^{(k)}||<\epsilon$ and $H$ rotated into $E^{(k)}$ from which, if one wants, one can read the eigenvalues. For small system sizes $L\leq 14$, on convergent samples, we have checked that the eigenvalues found in this way agree with those of exact diagonalization.

Let $v^{(k)}_S=||V_S^{(k)}||$, the flow is declared {\it resonant} at scale $k$ when the largest rotation angle that the generator would apply exceeds unity,
\begin{equation}
  \theta_k \;=\; \max_{S}\;\max_{\sigma}\,
  \left|\frac{v^{(k)}_S}{\partial_S E^{(k)}(\sigma)}\right| \;>\; 1 ,
  \label{eq:angle}
\end{equation}
the first maximum running over the off-diagonal Pauli strings of $V^{(k)}$ and the second over the background configurations $\sigma$ of the denominator's support ($\partial_S E$ is a diagonal operator in the $\sigma$ basis).

It is worth noting that on the converged samples, where by definition the Hamiltonian has been rendered local, the measured $\theta_k$ decay at least geometrically, possibly {\it a la} Newton-KAM $\theta_{k+1}=\theta_{k}^2$. So we could have substituted the constant $1$ in the RHS of \eqref{eq:angle} with a smaller number (say $1/2$) or even a decreasing function of $k$, without changing significantly the resulting set of converged states. 

This is sufficient to prove that the norm $||A_k||$ does not grow faster than $L$, but we need to prove also that the generators $A_k$ are (sums of) local operators. We check this later in the paper by decomposing $A_k$ over operators having support on an interval $I$ and checking that the norm of these operators decreases exponentially with the size of the interval.

\begin{table*}[t]
\caption{Correspondence between the objects of \cite{DRGHP} and those measured
here, with the status of each. The two schemes are not identical: DRGHP
eliminate at step $k$ only the diagrams of order $|g|<L_{k+1}$, whereas the
present flow eliminates the whole of $V^{(k)}$ at every step and has no analogue
of the scale separation $L_k=(1+\beta)^k$. We therefore do not identify the two
events, and no inequality between $P_{\rm conv}$ and $P_{\rm FNR}$ is claimed;
what is compared is the rate at which the Schrieffer--Wolff flow is obstructed,
estimated in the two cases by different means.}
\label{tab:correspondence}
\renewcommand{\arraystretch}{1.45}
\setlength{\tabcolsep}{8pt}
\begin{tabular}{|p{0.15\linewidth}|p{0.26\linewidth}|p{0.21\linewidth}|p{0.22\linewidth}|}
\hline
\textbf{Object} & \textbf{DRGHP} & \textbf{This work} & \textbf{Status} \\
\hline\hline
coupling &
$(\gamma/2)^{|I|}W_I$, all intervals, $\|W_I\|\le1$ &
$t\sum_i X_i$, single site only &
$t=\gamma/2$ from the $|I|=1$ term; a strict subclass \\
\hline
non-resonance, first kind &
$\NRI$, Eq.~(9.3): $\|1/D_{1,2}\|\le\varepsilon^{-|g|}$, on triple products of offset denominators &
$\theta_k=\max_S\max_\sigma$ $|c_S/\partial_SE|<1$, Eq.~(\ref{eq:angle}) &
same in spirit, not identical: one denominator per Pauli string, not a product per triad \\
\hline
non-resonance, second kind &
$\NRII$, Eq.~(9.4): $\|A^{(k)}(t)\|\le B_{II}(t)$ for non-crowded central diagrams &
not implemented &
no triads, orders or diagram factorials in the engine; not tested \\
\hline
consequence of the above &
Prop.~13: $\|V^{(k)}_I\|,\|A^{(k+1)}_I\| \le(\gamma/\varepsilon\delta)^{\max\{|I|,\beta L_k\}}$ &
$\Lambda(m)=\max_{|I|=m}$ $\sum_{{\rm supp}=I}\|X_Sf\|$ &
exponential decay in $m$ and normalization confirmed \\
\hline
scale index &
$L_k=(1+\beta)^k$; only $|g|<L_{k+1}$ removed at step $k$ &
all of $V^{(k)}$ removed at every step &
not in one-to-one correspondence \\
\hline
measured quantity &
$P_{\rm FNR}\ge e^{-\gamma^{c'}L}$, Theorem 1 &
$P_{\rm conv}$, at fixed $\epsilon$, $N_{\rm LOC}$, $D_{\rm max}$, $M_{\rm max}$ &
two estimates of the obstruction rate, not nested events \\
\hline
\end{tabular}
\end{table*}

Now, DRGHP proof, under two sets of conditions, which they call $\NRI$ and $\NRII$, that the SW flow converges. Condition $\NRI$ is about non-resonances and it is akin our (\ref{eq:angle}), while $\NRII$ is about the decay of amplitudes of $A^{(k)}, V^{(k)}$ decomposed over Pauli strings of a certain size. The process of construction of $A_k$ in DRGHP, leading to condition $\NRI$ however is not exactly the same as ours. At step $k$ they do eliminate operators whose support exceeds a given $L_k=(1+\beta)^k\simeq 2^k$, while we eliminate leading order, {\it all} off-diagonal operators and one can check that they grow at most to span $2^k$ sites. So the two processess are not, strictly speaking one-to-one and this means our resonance events are not a proper subset of theirs and {\it vice versa}. However both sets of {\it converged} events are contained in the proper set of samples on which the SW flow is convergent, so they both act as bound to the physical construction of the local unitary. 

While we test for (\ref{eq:angle}), we were not able not test for the second condition so we check, on the converged population, instead of $\NRII$, the necessary condition in Proposition 13 of DRGHP which follows from it. This we do later in the paper.

Since we have the above 4 conditions before getting to (\ref{eq:angle}) we need to make sure that the chosen parameters $\epsilon, N_{\rm LOC}, D_{\rm max}$ and $M_{\rm max}$ are chosen so that our numerical estimate of the convergence probability $P_{\rm conv}$, the fraction of samples which can be succesfully diagonalized using SW with a resulting local $A_k$ is tight.

We need to declare the nature of the numerical limitations. 

Condition (\ref{eq:angle}) and limitation number 4 on $N_{\rm LOC}$ make $P_{\rm conv}$ increases with increasing the maximum value of $\theta$ and increasing $N_{\rm LOC}$. Condition 1 and 2 instead work in the opposite direction $P_{\rm conv}=\lim_{\epsilon\to 0}P_{\epsilon}$ and $P_{\epsilon}$ decreases, possibly monotonically, with $\epsilon$. 

We tested the dependence on $\epsilon$ directly with the production code, by rerunning at $\epsilon=10^{-7}$ (with the pairwise threshold inside the commutator lowered to $0.1\,\epsilon$) a stratified sample of $106$ disorder realizations at $L=20,24$ and $t=0.04,0.045$ that reach step $3$ at $\epsilon=10^{-6}$. The strata are the outcome at $\epsilon=10^{-6}$ (converged, resonant at $k=3$, resonant at $k\geq4$), whether the sample needed a larger $N_{\rm LOC}$, $L$ and $t$, with the converged and the $k=3$ samples oversampled; the results are weighted by the sizes of the strata, which together contain $6830$ samples. Lowering $\epsilon$ from $10^{-6}$ to $10^{-7}$ raises the fraction of these samples that resonate at step $3$ from $14.4\%$ to $17.7\%$, while $1.3\%$ resonate already at step $2$. The increase comes mostly from samples that converge at $\epsilon=10^{-6}$: $7$ of the $68$ converged samples of the test resonate at step $3$ at $\epsilon=10^{-7}$; conversely, $8$ of the $27$ sampled step-$3$ resonances pass step $3$ at $\epsilon=10^{-7}$. The rates $a_{k\geq3}$ reported below, obtained at $\epsilon=10^{-6}$, are therefore likely to be underestimated; for $a_3$ at $t=0.04$--$0.045$ and $L=20$--$24$ the fraction resonating at step $3$ increases by about $25\%$. We have not measured the corresponding effect on $a_4$. A production run at $\epsilon=10^{-7}$ costs at least an order of magnitude more per sample and is beyond the present work.

Analogously for $D_{\rm max}$. Increasing $D_{\rm max}$ lowers the probability of convergence, since it adds more possibly resonating operator strings. In our production runs no weight at all is discarded by the diameter cutoff in $92$--$98\%$ of the converged samples (the fraction decreasing with $t$ over $t=0.025$--$0.045$), and the discarded weight exceeds $\epsilon$ in fewer than $0.1\%$ of the samples. The denominator cutoff $M_{\rm max}$ truncates a denominator in at most $0.3\%$ of the converged samples.

With all this in mind, we are now able to define the probability of convergence, broken down by the scale $k$ at which it has broken (surviving all the $k-1$ iterations before that):
\begin{equation}
    P_{\rm conv}=\prod_{k\geq 1}S_k(L),
\end{equation}
short-writing $S_k(L)=\mathbb{P}\!\left(\text{survive step }k\,\middle|\,\text{survived through step }k-1\right)$. The locality of the condition of resonance allows us to write
\begin{equation}
    S_k=A_k e^{-a_k(t) L},
\end{equation}
where $A_k(t,L)\simeq 1$ (confirming this is a Poisson rare events counting). The statement of {\it at least} exponentially small $P_{\rm conv}$ is a statement of convergence of of the series 
\begin{equation}
    a(t)=\sum_{k\geq 1}a_k(t).
\end{equation}
Physically, $a_k$ is the rate {\it per spin} of the existence of a step-$k$ resonance. This means, the probability that a spin belongs to a, possibly very large, resonance created only after step $k$, and not before. Computing this in perturbation theory is a highly complicated task, since the number of independent events that can give rise to a resonance is a difficult object to estimate at any step $k$. Bounding it is a main feat of DRGHP. The fact that we can define it and measure it numerically, although with a complex numerical algorithm, is the central result of this paper. 

We also checked the Proposition 13 of the DRGHP paper, decomposing the generator and the remaining off-diagonal terms into Pauli strings and checking the weight on strings of size $m$. This is necessary to ensure locality of the ensuing unitary transformation, which is the strongest definition of MBL (in the sense that it is a sufficient but not necessary condition for the vanishing of conductance, which was the target of Anderson \cite{Anderson58} and BAA \cite{basko2006metal}). We observe, on converged samples, exponential decay of the weights with $m$, which is again in confirmation of DRGHP. 

\section{Numerical estimates of convergence probability}
\label{sec:numerics}

Before running the SW iterations it is worth recognizing that the first coefficient $a_1$ can be computed analytically. In fact, the probability of resonance at this point is just the probability that a single spin flip is resonant. This can be computed easily, knowing the distributions of the {\it iid} variables $h_i,J_i$ and the value of $t$. The result (see Sec.~\ref{sec:analytic} for the calculation) is
\begin{equation}
    a_1(t)=\frac{3}{2}t.
\end{equation}
This observation has a double benefit. It gives us an analytic handle on $a_1$ but also allows us to pre-select the samples which {\it do not} have a $k=1$ resonance, therefore saving time and computational resources to concentrate on $a_{k\geq 2}$. Therefore we report the numerics on the {\it conditioned ensemble} of samples lacking the single-spin resonance.

\begin{figure}[!htbp]
    \centering
    \includegraphics[width=0.85\linewidth]{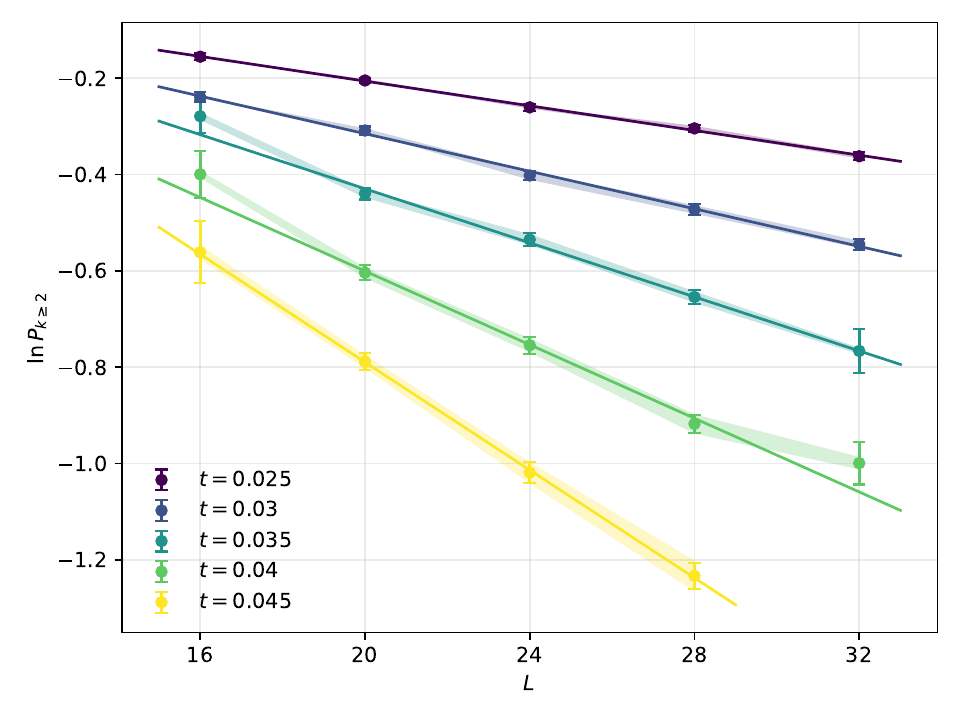}
    \caption{The measured convergence probability $P_{k\geq2}$ of the SW iterations, $L=16,20,\dots,32$, on the conditioned ensemble without single-spin ($k=1$) resonances. Lines are weighted least-squares fits $\ln P_{k\geq2}=c-a_{k\geq2}L$. The bands, barely visible, bracket the few samples still unresolved by the density guard, counted as all diverging (lower edge) or all converging (upper edge); error bars are binomial.}
    \label{fig:lnP_L}
\end{figure}

\begin{figure}[!htbp]
    \centering
    \includegraphics[width=0.85\linewidth]{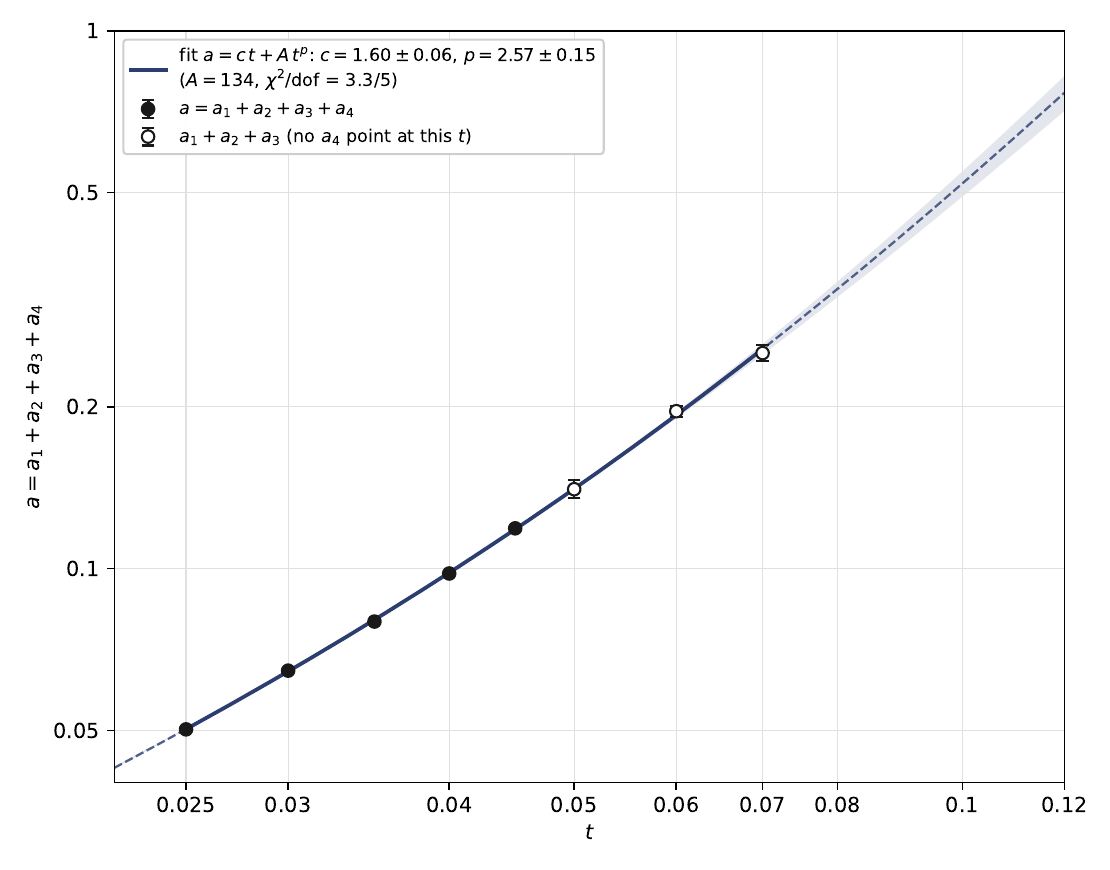}
    \caption{The total rate for any event $k\geq 1$ as a function of $t$. The increase is faster than $t$, but there is no sign of divergence. Some hint of a radius of convergence can be seen by separating the $a$ into $a_k$'s.}
    \label{fig:ak2}
\end{figure}

\begin{figure}[!htbp]
    \centering
    \includegraphics[width=0.95\linewidth]{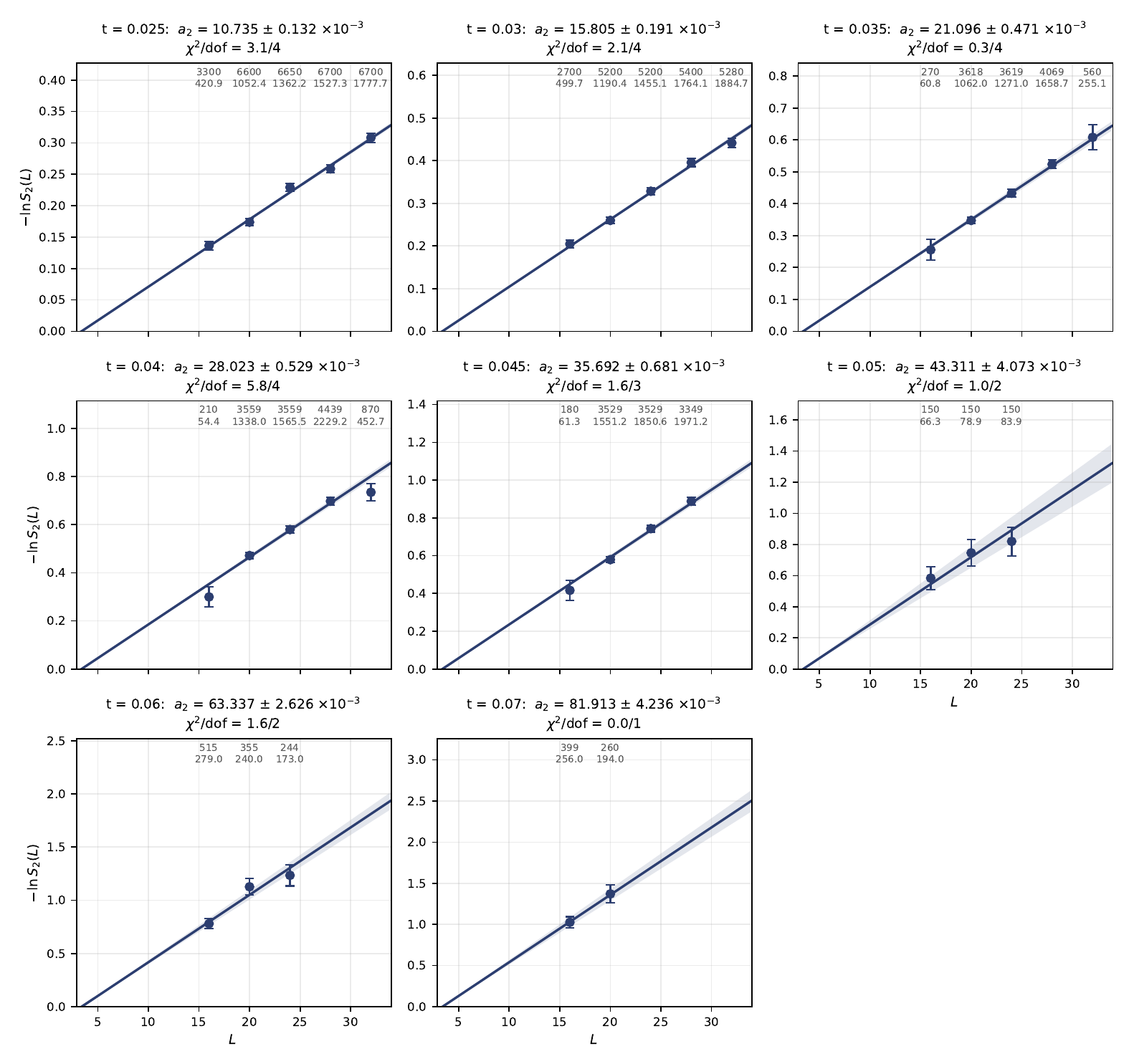}
    \caption{Extraction of $a_2(t)$ from the size dependence of the survival probability of step $k=2$. Each panel is one value of $t$ and shows $-\ln S_2(L)$ for $L=16,\dots,32$ (fewer sizes at $t\geq 0.045$). The line is the binomial maximum-likelihood fit $-\ln S_2=a_2(L-L_{0,2})$ with the offset $L_{0,2}=3.4$ common to all $t$, and the shaded band its $68\%$ statistical error from a bootstrap over disorder realizations. The fitted $a_2$ and the $\chi^2$/dof are in the panel titles. The numbers on top of each point are the samples at risk (having survived step $1$) and the failures at step $2$; the point is $-\ln(1-\text{failures}/\text{at risk})$. The failures are not integers: the samples still unresolved by the density guard are counted as half converging and half diverging, and the samples that were not rerun with the production code enter with weights taken from the rerun samples of the same class and $t$ (for $k=4$, Fig.~\ref{fig:a4_Lfit}, the failures are also rescaled by the bulk fraction). The grey bars, when visible, show the range spanned by counting the still unresolved samples as all converging or all diverging. The number of samples differs between points because the data at different $(t,L)$ come from different runs; at $t\geq0.035$ the sizes $L=16$ and $L=32$ have the fewest samples.}
    \label{fig:a2_Lfit}
\end{figure}

\begin{figure}[!htbp]
    \centering
    \includegraphics[width=0.95\linewidth]{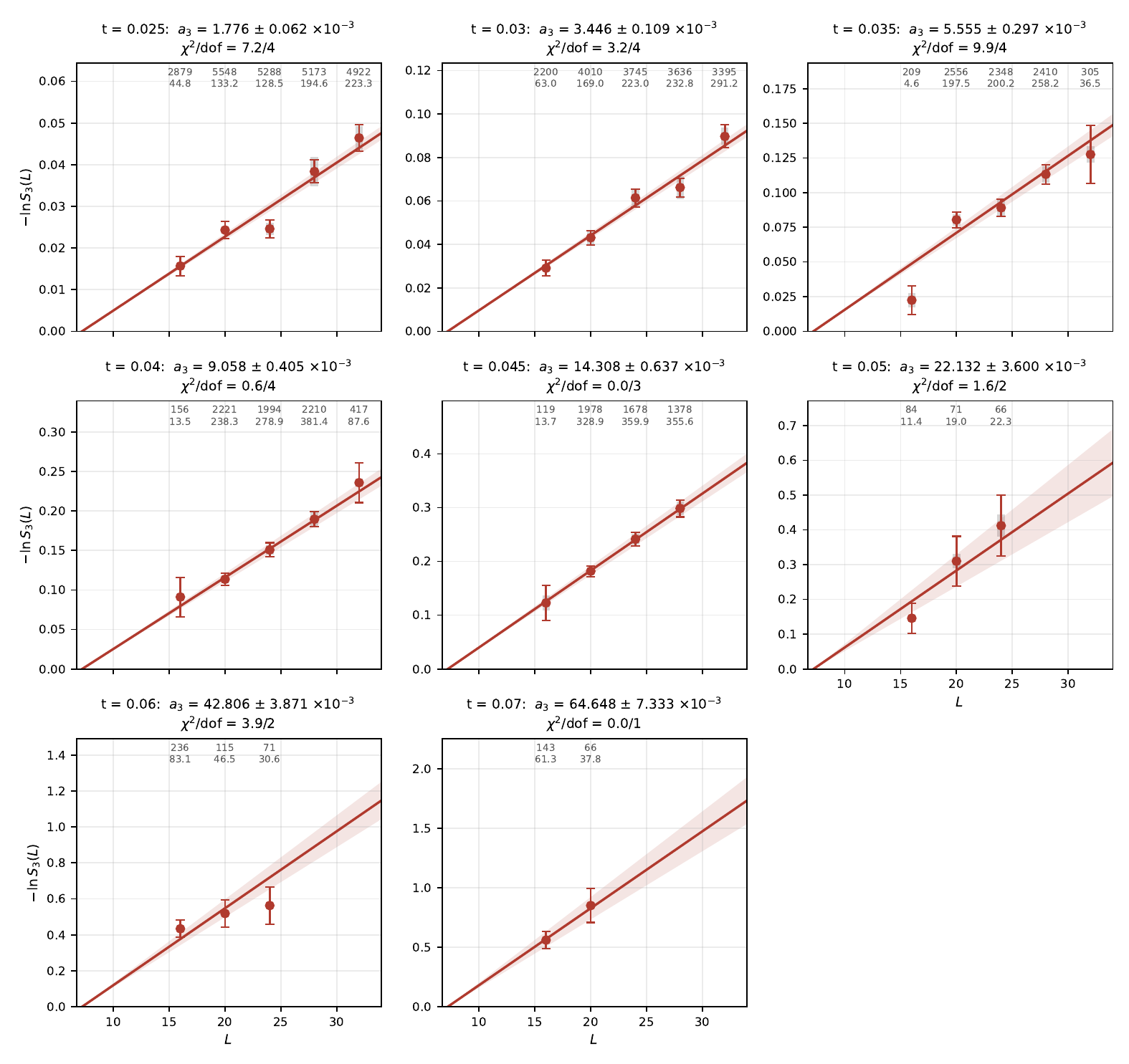}
    \caption{The same as Fig.~\ref{fig:a2_Lfit} for the third step: $-\ln S_3(L)$ on the samples which survived steps $1$ and $2$, with the fit $-\ln S_3=a_3(L-L_{0,3})$ and the common offset $L_{0,3}=7.2$.}
    \label{fig:a3_Lfit}
\end{figure}

\begin{figure}[!htbp]
    \centering
    \includegraphics[width=0.95\linewidth]{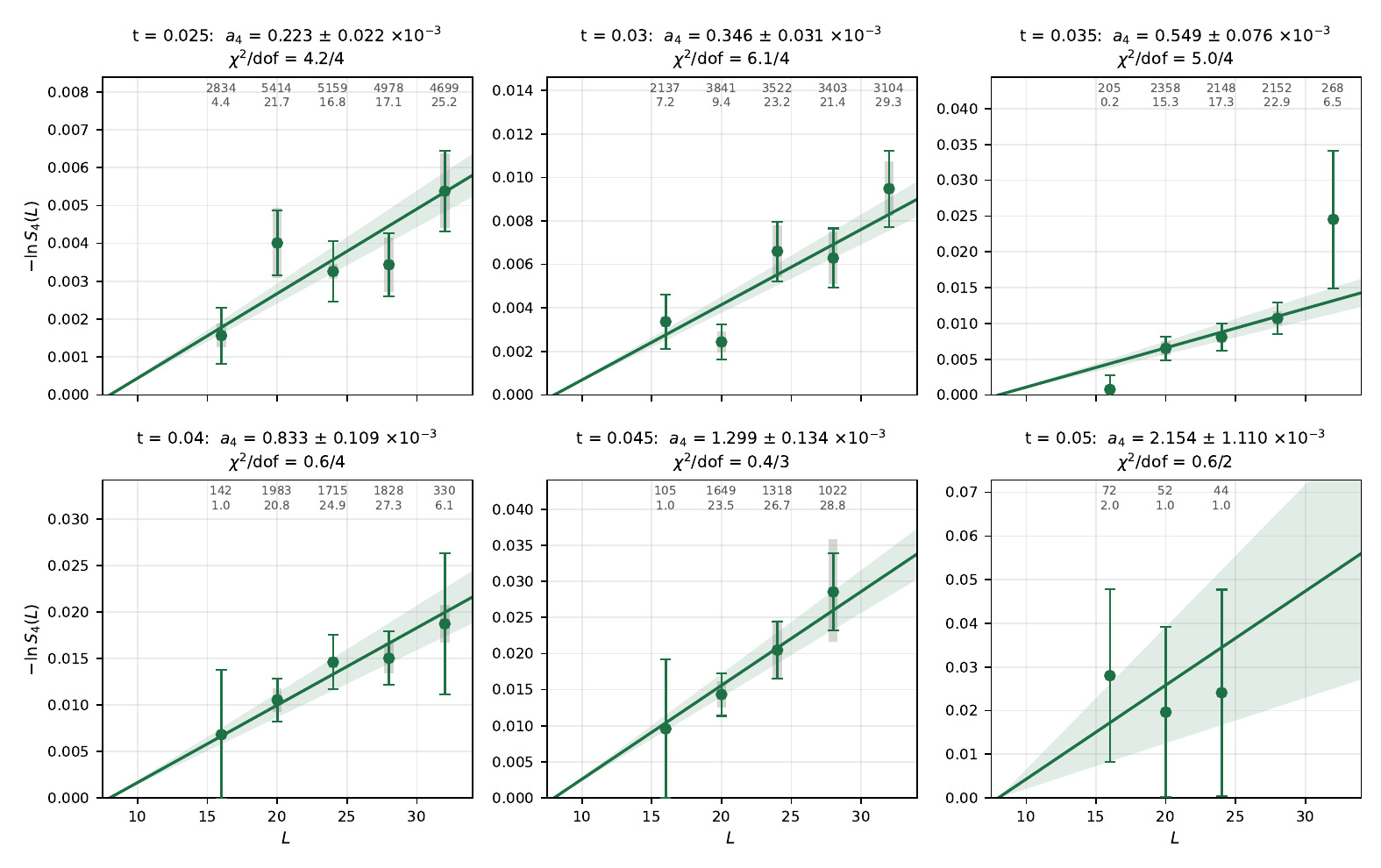}
    \caption{The same as Fig.~\ref{fig:a2_Lfit} for the fourth step. Here we count only the step-$4$ resonances in the bulk, at least $4$ sites away from each edge, so the exposed length is $L-8$ and the line is $-\ln S_4=a_4(L-8)$. The failures are rescaled by the measured bulk fraction of step-$4$ resonances at each $L$. With at most about $30$ failures per point, $a_4$ has the largest relative errors. At $t=0.05$ the three points carry one or two failures each, and this value is not used in Fig.~\ref{fig:ak_t}.}
    \label{fig:a4_Lfit}
\end{figure}

\begin{figure}[!htbp]
    \centering
    \includegraphics[width=0.85\linewidth]{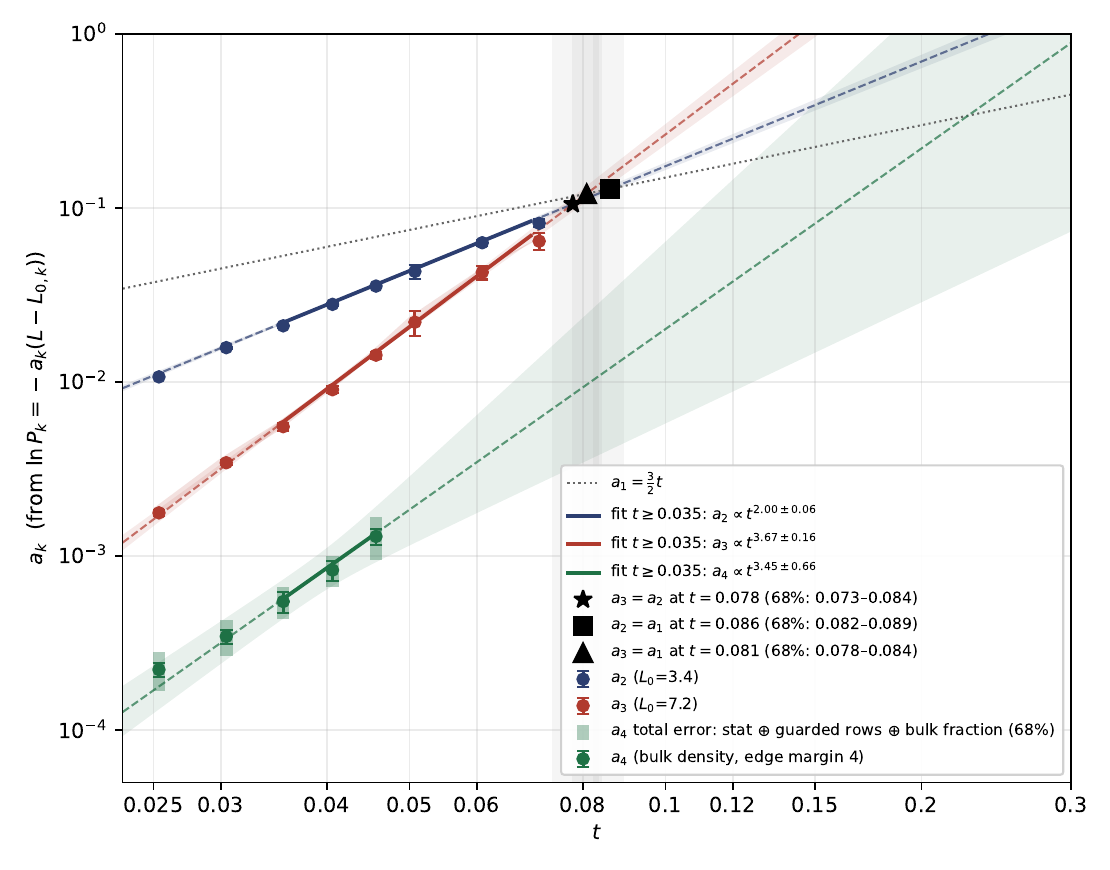}
    \caption{The rates $a_2,a_3,a_4$ from the data, obtained as in Figs.~\ref{fig:a2_Lfit}--\ref{fig:a4_Lfit}. The analytic result $a_1=(3/2)t$ is the dotted line. The lines are power-law fits for $t\geq 0.035$ (solid over the fitted range, dashed outside), with their $68\%$ bands: $a_2\propto t^{2.00\pm0.06}$, $a_3\propto t^{3.67\pm0.16}$ and $a_4\propto t^{3.45\pm0.66}$. The shaded boxes on $a_4$ are its total error (statistical, unresolved samples and bulk fraction). The $a_4$ fit rests on three points and its extrapolation is very uncertain. The three crossings $a_3=a_2$ at $t=0.078$ ($68\%$: $0.074$--$0.083$), $a_3=a_1$ at $t=0.081$ ($0.078$--$0.084$) and $a_2=a_1$ at $t=0.086$ ($0.082$--$0.089$) occur at very close points. This is suggestive of the failure of the perturbative regime and a possible divergence of the series $\sum_{k\geq 1}a_k(t)$ in that bracket.}
    \label{fig:ak_t}
\end{figure}

\begin{figure}
    \centering
    \includegraphics[width=0.95\linewidth]{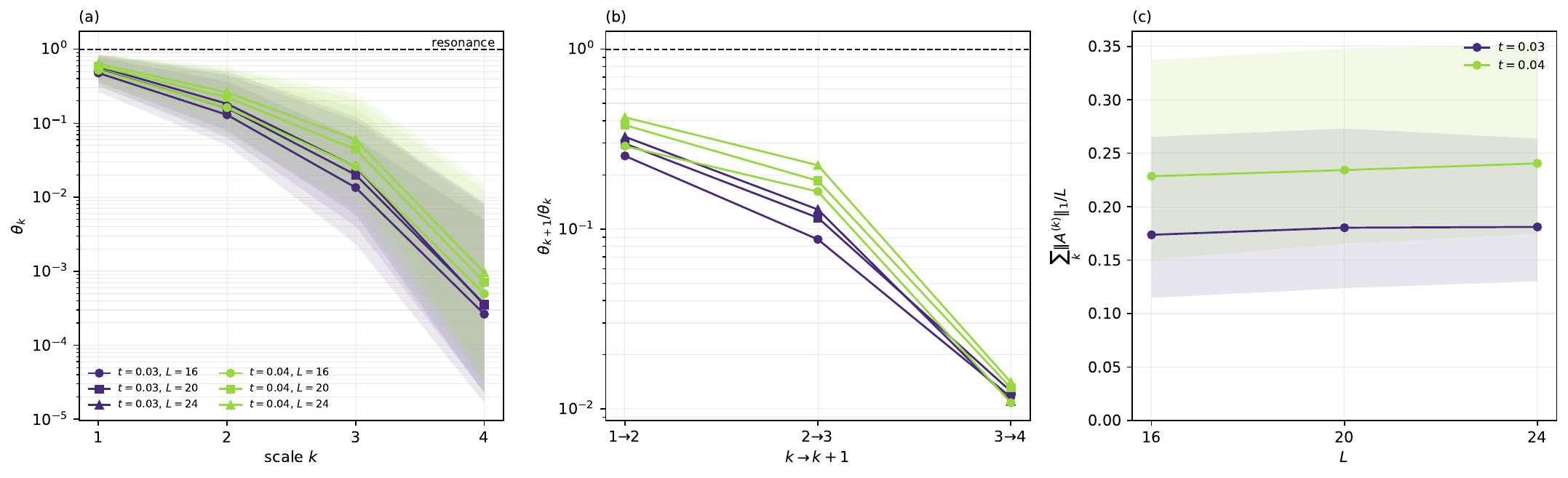}
    \caption{The rotation angles $\theta_k$ at step $k$ for $t=0.03,0.04$ and $L=16,20,24$ on about $12\,700$ converged samples (medians, with $16$--$84\%$ bands). {\it Left panel,} their typical, median values in log-linear plot; {\it Central panel} their ratio which shows that the typical value of $\theta_k$ seems to decay faster than exponential; {\it Right panel} An upper bound $||A_k||_1$ norm to check that the normalization of the rotation generator $A_k$ grows not faster than linearly in $L$. The operator norm $||A_k||\leq ||A_k||_1$ but it is quite expensive to compute since it requires exact diagonalization of $A_k$.}
    \label{fig:placeholder}
\end{figure}

The first result is that the decay of $P_{\rm conv}(L)$ is perfectly exponential, see Fig.\ref{fig:lnP_L}. The rate of decay $a(t)$ is an increasing, convex function of $t$ in Fig.\ref{fig:ak2}. There is no sign of divergence of $a(t)$, signalling that our range $t\leq 0.07$ is far from the critical $t_c$. An estimate of $t_c$ is therefore impossible from this quantity. The exponential decay of $P$ is the main result of DRGHP and what implies the existence of rare, localized regions which are sufficient to stop diffusion. Quantitatively, the rate measured is much larger than the guaranteed rate of DRGHP. At $t=0.03$ ($\gamma=2t=0.06$) and $L=32$ the constants of \cite{DRGHP} give a guaranteed rate $\gamma^{c'}=0.9987$ per site, since (4.3) and (4.8) fix $a(\beta)<4.58\times10^{-4}$, and hence $P_{\rm FNR}\ge e^{-0.9987\times32}=1.3\times10^{-14}$, whereas the measured rate $a=\tfrac32t+a_{k\ge2}=0.0646$ gives $P_{\rm conv}=0.127$. The two differ by thirteen orders of magnitude in probability and by a factor $15$ in the rate. Because the exponent $c'$ is very small, $\gamma^{c'}\simeq 1$, so the guaranteed rate comes up at one resonance per site while the true rate vanishes linearly in $\gamma$.

The separate rates $a_k(t)$ are obtained from the size dependence of the step survival probabilities $S_k(L)$, shown in Figs.~\ref{fig:a2_Lfit}, \ref{fig:a3_Lfit} and \ref{fig:a4_Lfit} for $k=2,3,4$. At every $t$ and for every $k$, $-\ln S_k(L)$ is linear in $L$ within the errors over the range $L=16,\dots,32$, as expected from the rare-event form $S_k=A_k e^{-a_k L}$. We write $A_k=e^{a_kL_{0,k}}$, so that $-\ln S_k=a_k(L-L_{0,k})$, and fit by binomial maximum likelihood the number of failures at step $k$ among the samples that survived the previous $k-1$ steps. The offset $L_{0,k}$ accounts for the boundaries: a resonance at step $k$ involves a region of several sites, which has fewer positions available near the edges of an open chain. We take one offset per $k$, common to all $t$, from the profile likelihood: $L_{0,2}=3.4$ (68\% range $2.9$--$3.9$) and $L_{0,3}=7.2$ ($6.3$--$8.0$); $L_{0,k}=0$ is excluded at more than $5$ standard deviations for both. For $k=4$ the events are too few to fit an offset, and we instead restrict the count to bulk resonances, at least $4$ sites from each edge, with exposed length $L-8$ (Fig.~\ref{fig:a4_Lfit}). With these offsets $A_k\simeq 1$ at small $t$: $e^{a_2L_{0,2}}=1.04$ and $e^{a_3L_{0,3}}=1.01$ at $t=0.025$.

The fits have $\chi^2$/dof of order one, with the largest values $5.8/4$ for $a_2$ ($t=0.04$), $9.9/4$ for $a_3$ ($t=0.035$, driven by the $L=16$ point) and $6.1/4$ for $a_4$ ($t=0.03$). Since the same disorder realization is used at different $L$ (the shorter chain is the beginning of the longer one), the points at different $L$ are correlated; the error bands on $a_k$ therefore come from a bootstrap that resamples whole disorder realizations across all $L$ and $t$, and the $\chi^2$, which treats the points as independent, is only indicative. The relative statistical error is $1$--$2\%$ for $a_2$, $3$--$6\%$ for $a_3$ and $9$--$14\%$ for $a_4$ at $t\le0.045$, which are the values reported in Fig.~\ref{fig:ak_t}. For $a_4$ the residual uncertainty from samples whose classification is still open is comparable to the statistical one and is included in the total error bars of Fig.~\ref{fig:ak_t}. At $t=0.05$ the $a_4$ estimate rests on about four failures in total (Fig.~\ref{fig:a4_Lfit}, last panel) and is not shown in Fig.~\ref{fig:ak_t}.

Some words are now in order regarding the purported transition at $t_c\simeq 0.08$. Of course, estimating the divergence of a series by the ratio of the first 3 terms is not good practice. However, the possibility to split the ratios into separate $S_k$'s and study them separately is a powerful tool. There are two reasons for this. First, by looking at the total rate $a(t)$ there is no sign of divergence, $a(t)$ growing slightly faster than $t$. The region of exponentially decreasing probability, the {\it Griffiths region}, looking at $a(t)$ seems to extend for all $t\in\mathbb{R}$. This cannot possibly be the case, and one should not believe in the data observed on a small region of parameters to be representative of the whole phase diagram. Similar conceptual missteps {\it mutatis mutandis} have led people to make statements that the MBL region {\it does not exist} in spin chains. Breaking down $a(t)=\sum_{k}a_k(t)$ gives us a much more powerful tool. In fact, the second observation is that the physical interpretation of the statement $a_{k+1}(t)/a_{k}(t)=1$ at $t=t_c$ is clear. The problem becomes scale free: the probability to have a resonance at scale $k$ is independent of $k$. This is a signature of the starting of an {\it avalanche} process \cite{gopalakrishnan2019instability,morningstar2022avalanches,colmenarez2024ergodic}, which continues to the largest $k$. Since the size of the regions involved at step $k$ grow exponentially with $k$ (one $X_i$ spin flip at $k=1$, two adjacent spins $X_i X_{i+1}$ at $k=2$ and 4 adjacent spins at $k=3$) the largest $k_{\rm max}(L)\sim \ln L$. This means that, at $t_c$ the probability is smaller than any exponential
\begin{equation}
    P_{\rm conv} (t>t_c,L)\sim e^{-b L\ln L}.
\end{equation}
If this were indeed the behavior of the full SW flow, such a small rare-region probability would not be sufficient to act as a bottleneck to stop the diffusivity and transport becomes again diffusive. In a later Section we introduce a toy model based on the forward scattering approximation \cite{ros2015integrals,pietracaprina2016forward} which qualitatively explains the whole situation.

\section{Locality of the $V_k$'s and $A_k$'s and further points of contact with DRGHP}
\label{sec:prop13}

Assuming the SW flow converges, one should check that the generators $A_k$ are sum of local operators themselves. In this way, the unitary $U$ will be product of a small set of local unitaries. This implies the existence of LIOMs, and by the argument of Agarwal {\it et al} \cite{agarwal2017rare, schulz2020phenomenology} leads to subdiffusion in the thermodynamic limit.\footnote{To obtain subdiffusion one must also require an extra condition between the rate $a$ of the convergence probability for the flow, and the value of the resistance $R\sim e^{\eta L}$ of such localized samples, which, in the range explored, is verified.} 

The convergence itself does not imply the locality of $A_k$ and this is exactly what we want to check now. In DRGHP the locality of $A_k, V_k$ (in principle we need only $A_k$ but it is good to check $V_k$ as well) is guaranteed by their $\NRII$ condition.

As said before, $\NRII$ is a bound on the decay of the operators $A,V$ supported on a given {\it triad} of a given size. This second bound is important in their proof, since it is not inductive (such a lack of induction was already pointed out by Imbrie \cite{Imbrie16}) and needs a separate discussion. A direct check of $\NRII$ (with triads and all that) for us is impossible within the current implementation of the numerics. The closest thing we can check is a consequence of that bound, the locality of $A^{(k)}$, as stated in Proposition 13 of DRGHP. 

Proposition~13 of \cite{DRGHP} bounds the local parts of the perturbation and of
the generator on the full non-resonance event,
\begin{equation}
  \big\|V^{(k)}_I\big\|,\ \big\|A^{(k+1)}_I\big\|
  \;\le\; \Big(\frac{\gamma}{\varepsilon\delta}\Big)^{\max\{|I|,\,\beta L_k\}},
  \label{eq:prop13}
\end{equation}
and is the natural target for a numerical check: unlike the non-resonance
condition $\NRII$ of their (9.4), which is stated per triad and involves the
diagram order, bare order and diagram factorial, \eqref{eq:prop13} refers only
to the part of an operator supported on an interval. 

For both $V^{(k)}, A^{(k)}$ we employ the decomposition in terms of $X_S$ monomials as before:
\begin{eqnarray}
    V^{(k)}&=&\sum_S X_Sf^{V}_S(Z),\\
    A^{(k)}&=&\sum_S X_Sf^{A}_S(Z),
\end{eqnarray}
where $f^{A,V}$ contain a sum over $Z-$ monomials and identities.

We therefore measure, for each interval width $m$, the largest total weight of $X$-monomials whose support
spans exactly that width (analogously for $V,A$ -- we do not write the superscript explicitly). 

\begin{figure}
    \centering
    \includegraphics[width=0.95\linewidth]{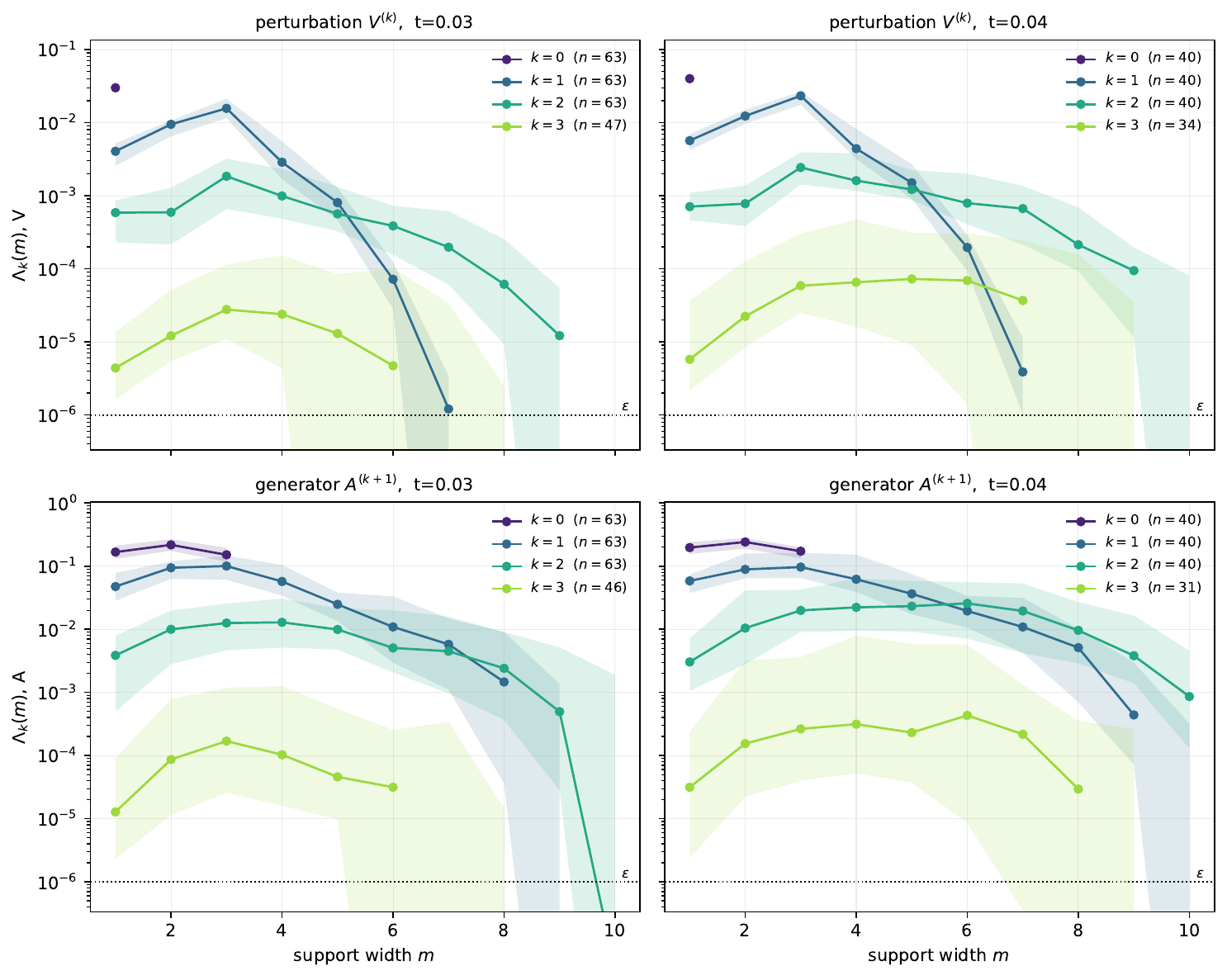}
    \caption{Check of the locality of $V^{(k)}$ (top) and $A^{(k+1)}$ (bottom) required by
    Proposition~13 of \cite{DRGHP}: the profile $\Lambda_k(m)$ of \eqref{eq:lambda} against the
    support width $m$, at $L=24$, $t=0.03$ (left) and $t=0.04$ (right), with the production code
    ($\varepsilon=10^{-6}$, $D_{\max}=10$). Points: median over the converged samples
    whose operator at scale $k$ is non-empty ($n$ in the legend), zeros included; bands: 25--75\%
    quantiles. A point is omitted where the median vanishes. Strings are cut at diameter
    $D_{\max}$; the dotted line marks $\varepsilon$.}
    \label{fig:Lambdas}
\end{figure}

We then define:
\begin{equation}
  \Lambda_k(m) \;=\; \max_{|I|=m}\ \sum_{\substack{X_Sf_S\ :\ \mathrm{supp}=I}}
                     \big\|X_Sf_S\big\| ,
  \label{eq:lambda}
\end{equation}
with $\|X_Sf\|=\max_\sigma|f(\sigma)|$ evaluated by inverse Walsh transform of
the retained strings, cf.\ (4.16) of \cite{DRGHP}. The results are in Figure \ref{fig:Lambdas}.

The decay of the $\Lambda_k(m)$ with $m$ is evident in Figure \ref{fig:Lambdas}, for both $A,V$: it is exponential for $k\leq2$, while at $k=3$ the profile is flatter at intermediate $m$ (at $t=0.04$ the median of $V^{(3)}$ is flat over $m=3$--$6$) and decays at large $m$. Therefore, the unitary $U$ is a low-depth circuit of local rotations, and this guarantees the existence of LIOMs on the sample, in the most extreme definiton of MBL.

The exponential decay invokes a more quantitative comparison with DRGHP, however, as we see, this needs caution. Adopting their parameter choice $\varepsilon=\delta=\gamma^{a(\beta)}$ from their Eq.~(4.8), so that
$\gamma/\varepsilon\delta\simeq1.003\,\gamma$ at the couplings studied, the
comparison succeeds at the initial scale and fails at every later one. At $t=0.02$ and $L=8$ (production code, $\varepsilon=10^{-6}$; geometric means over
the 25 converged samples out of 30) we find $\Lambda_0(1)=2.00\times10^{-2}$; identifying $\gamma$
with $t$ (rather than with $2t$, as the coupling $(\gamma/2)^{|I|}$ of \eqref{eq:DRGHPmodel} would
give) and $m$ with $|I|$, the bound is $2.01\times10^{-2}$, an agreement to $0.4\%$; but at the next
scale $\Lambda_1(3)=3.8\times10^{-3}$ (sample range $0.9$--$15\times10^{-3}$) against
$8.1\times10^{-6}$, a factor of about $470$, and the discrepancy grows with $m$
($\Lambda_1(5)=9.3\times10^{-5}$ against $3.2\times10^{-9}$). (With the
coupling of \eqref{eq:DRGHPmodel}, $\gamma=2t$, the bounds double at $m=1$ and grow by $2^m$ in
general: $\Lambda_0(1)$ is then a factor $2$ below its bound, $4.01\times10^{-2}$, and $\Lambda_1(3)$
a factor $59$ above its bound, $6.5\times10^{-5}$.) The numbers at $\varepsilon=10^{-7}$ are the
same to the digits quoted.

We do not interpret this as a failure of \eqref{eq:prop13}, because there are sufficient differences between our two schemes that some quantities might be lost in translation. One candidate explanation for the discrepancy, that
$\Lambda_k(m)$ over-counts $\|V^{(k)}_I\|$ because an $\ell^1$ sum of Pauli
strings exceeds an operator norm, was tested and excluded: replacing the $\ell^1$
sum by the sup-norm of each reconstructed $X$-monomial changes the profile by
about one percent. A second, that DRGHP support convention
$\bar I=[\min I-1,\max I+1]$ shifts $m$ by two, accounts for the first scale but
not the second, and is inconsistent with the agreement at scale zero. The
explanation we consider most likely concerns the correspondence between scales.
The perturbation eliminated at step $k$ of their scheme is $V^{(k)}_{\rm per}$ of
their (6.10), which retains only diagrams of order $|g|<L_{k+1}$, whereas the
present engine constructs a generator for the whole of $V^{(k)}$ at every step
and has no analogue of the scale separation $L_k=(1+\beta)^k$. A given step of
our flow therefore need not correspond to a single scale of theirs, and applying
the floor $\beta L_k$ in the exponent of \eqref{eq:prop13} to our step-$k$ output
has no justification --- which would account both for the exact agreement at the
initial scale, where the floor is inactive, and for the discrepancy at the later
scales. A third possibility, that several of their diagrams share one $X$-support
and are summed together in \eqref{eq:lambda}, cannot be separated by any
string-level quantity and would require diagram labels to be propagated through
the flow. We note that establishing the correspondence is a prerequisite for any
quantitative comparison with the inductive bounds of \cite{DRGHP}, and that the
locality statement, which concerns the shape of $\Lambda_k(m)$ rather than its normalization, is unaffected by it.

In conclusion, the locality of $A_k,V_k$ is numerically proved in our work, but the rate is in disagreement with DRGHP. The reason why is not entirely clear.

\section{Analytical results and a toy model}
\label{sec:analytic}

In this Section we collect some, not many, unfortunately, analytic results we can obtain for the rates $a_k$. $a_1$ can be solved exactly, as mentioned before, and $a_2$ can be obtained in the {\it forward scattering approximation} \cite{pietracaprina2016forward} which should be valid for asymptotically small $t$. A toy model based on this approximation can be bootstrapped to all $a_k$'s and we can get some information on these rates (for example that $a_k\sim t^k\ln(1/t)$) and, once the multiplicity of the contributions it neglects is accounted for by a phenomenological constant $c$, it shows a transition at $t_c=2/(3c)$ with scale-invariant avalanches, $a_{k}(t_c)=a_{k+1}(t_c)$.

\subsection{Analytic expression for $a_1(t)$}

At the first scale, $V=t\sum_iX_i$ and $E=E^{(0)}$, so \eqref{eq:denominator}
gives for $v=X_i$
\begin{equation}
  \partial_S E(\sigma) = -2\,\sigma_i\big(h_i + J_{i-1}\sigma_{i-1}
  + J_i\sigma_{i+1}\big),
\end{equation}
and the angle \eqref{eq:angle} exceeds unity iff
\begin{equation}
  \big|h_i \pm J_{i-1} \pm J_i\big| < t/2
  \label{eq:app:k1res}
\end{equation}
for one of the four sign choices. With $h,J,J'$ i.i.d.\ uniform on $[-1,1]$, it is possible to compute the probability $ p(t)=P(\big|h_i + J_{i-1} + J_i\big| < t/2)$ and multiply the result by 4. Through elementary calculations one finds $p(t)=\frac{1}{96}(36t-t^3)\simeq \frac{3}{8}t$ so, to leading order in $t$, the failure rate per bulk site is
\begin{equation}
  a_1(t) = 4\tfrac{3}{8}\,t = \tfrac{3}{2}\,t.
  \label{eq:a1}
\end{equation}
This expression was checked against numerics for the whole range of $t$ explored, and it is used in all the results of the paper.

Because \eqref{eq:a1} is exact and \emph{checkable in closed form for a given disorder realization} -- it is $4L$ inequalities, one per site per sign choice, requiring no Pauli string -- the first scale need not be simulated at all. This saves time but also gives an analytic form which furnishes the first crossing point $a_1(t)=a_2(t)$ with considerable accuracy since no error is associated to $a_1(t)$.

The rotation generator is the anti-hermitian matrix
\begin{equation}
    A^{(1)}=\sum_i X_ia_i(Z),
\end{equation}
and
\begin{equation}
    a_i(Z)=\frac{t}{\partial_i E(Z)},
\end{equation}
and its expression will be used in the next subsection.

\subsection{Higher steps failure rates $a_{k\geq 2}(t)$}

A similar calculation is impossible for $a_{2}(t)$ and higher $k$'s, since the first step renormalizes the energies to $O(t)$. However, in the spirit of the forward scattering approximation \cite{ros2015integrals,pietracaprina2016forward}, if one neglects such renormalization one can compute the lowest order in $t$ of $a_2(t)$.

The leading new off-diagonal terms come from $[A^{(1)},V^{(0)}]$. $A^{(1)}$ is chosen to eliminate the 0-th order term: $[A^{(1)},E^{(0)}]=-V^{(0)}$, but the higher terms in the BCH formula generate new off-diagonal couplings, albeit of higher order in $t$
\begin{equation}
  e^{A^{(1)}}He^{-A^{(1)}} = E + V - V + [A^{(1)},V] + \tfrac12\big(-[A^{(1)},V]\big) + O(A^{(1)\,2}V)
              = E + \tfrac12[A^{(1)},V] + O(A^{(1)\,2}V).
  \label{eq:app:half}
\end{equation}
In general, using $\mathrm{ad}_A^nE=-\mathrm{ad}_A^{n-1}V$, the coefficient of $\mathrm{ad}_A^nV$ is $1/n!-1/(n+1)!=n/(n+1)!$, whose leading value is $1/2$.
These new terms can be written as
\begin{equation}
    V^{(1)}=\tfrac12\sum_{ij}X_iX_j\left(a_i^{(j)}(Z)-a_i(Z)\right)
\end{equation}
where we have used $X_jf(Z)=f(\dots,-Z_j,\dots)X_j$,
\begin{equation}
  \big[X_i a_i(Z),\,tX_j\big] \;=\; t\,X_iX_j\,\big(a_i^{(j)}-a_i\big),
\end{equation}
where $a_i^{(j)}$ denotes $a_i$ with $\sigma_j\to-\sigma_j$. Three cases arise.
For $j=i$ the operator $X_iX_i=\mathbb{I}$ and the term is diagonal --- it is the
renormalization of the energy, and contributes to $E^{(1)}$, not to $V^{(1)}$. This is the term we are going to ignore altogether in the following (but it is kept and computed in the code).

For $|i-j|>1$ the function $a_i$ does not depend on $\sigma_j$, and the commutator vanishes identically. Only $j=i\pm1$ survives, and produces the two-site flip $X_iX_{i+1}$. The proliferation at this
order is one new object per bond.

Collecting both orderings on the bond $(i,i+1)$ and writing
\begin{equation}
  u \;=\; h_i + J_{i-1}\sigma_{i-1},
  \qquad
  v \;=\; h_{i+1} + J_{i+1}\sigma_{i+2},
  \qquad
  J \;\equiv\; J_i ,
\end{equation}
the coefficient of $X_iX_{i+1}$ in the expansion of $V^{(1)}=\sum_i X_i X_{i+1}f^{V^{(1)}}_i(Z)$ is
\begin{equation}
  f^{V^{(1)}}_i(\sigma) \;=\; t\big[(a_i^{(i+1)}-a_i)+(a_{i+1}^{(i)}-a_{i+1})\big]
            \;=\; -\,t^2 J\,\sigma_i\sigma_{i+1}
              \left[\frac{1}{u^2-J^2}+\frac{1}{v^2-J^2}\right],
  \label{eq:app:c}
\end{equation}
where we have used
\begin{eqnarray}
    a^{(i+1)}_i&=&-\frac{t}{2\sigma_i}\left(\frac{1}{u-J\sigma_{i+1}}-\frac{1}{u+J\sigma_{i+1}}\right)\nonumber\\
    &=&2Jt\frac{\sigma_i\sigma_{i+1}}{u^2-J^2}.
\end{eqnarray}
Note that $|v|$ depends on the background only through $\sigma_{i-1}$ and $\sigma_{i+2}$.

Now for the denominator. For the $X$-monomial $X_iX_{i+1}$ the active set is $S=\{i,i+1\}$, and by
\eqref{eq:denominator} the contributing $Z$-strings are those with \emph{odd}
overlap with $S$. Of the five diagonal terms touching the bond, four qualify ---
$h_iZ_i$, $h_{i+1}Z_{i+1}$, $J_{i-1}Z_{i-1}Z_i$, $J_{i+1}Z_{i+1}Z_{i+2}$ --- while
$J_iZ_iZ_{i+1}$ has even overlap and \emph{drops out}. Hence
\begin{equation}
  \partial_{S}E^{(1)}(\sigma)
  \;=\; -2\big(\sigma_i u + \sigma_{i+1} v\big) \;+\; O(t^2),
  \label{eq:app:dE2}
\end{equation}
the $O(t^2)$ being the energy renormalization generated above, which, as said, will be neglected in the spirit of the forward scattering approximation. 

Two features differ from the first scale: the sum runs over four disorder
variables rather than three, and it organizes into the same combinations $u,v$
that appear in the amplitude \eqref{eq:app:c}. The two are therefore correlated.

Armed with numerator ad denominator, we can compute now the probability that a resonance at the second scale occurs. That is, that the rotation angle exceeds unity,
\begin{equation}
  \big|\sigma_i u + \sigma_{i+1} v\big| \;<\; R,
  \qquad
  R \;\equiv\; \frac{|v_i|}{2}
    \;=\; \frac{t^2|J|}{2}\left|\frac{1}{u^2-J^2}+\frac{1}{v^2-J^2}\right| .
  \label{eq:app:res}
\end{equation}
The background enters through four spins: $\sigma_{i-1}$ and $\sigma_{i+2}$
select the branch of $(u,v)$, while $\sigma_i,\sigma_{i+1}$ enter only through
$|\pm u\pm v|$, which takes the two values $|u+v|$ and $|u-v|$, each twice.
There are therefore
\begin{equation}
  2\times2\ \text{(branches)}\ \times\ 2\ \text{(values)} \;=\; 8
\end{equation}
distinct conditions per bond, and one bond per site. Using this counting in a probability of course assumes there is no correlation between the conditions -- which is an approximation.

For a fixed branch, $u=h_i\pm J_{i-1}$ and $v=h_{i+1}\pm J_{i+1}$ are independent
sums of two variables uniform on $[-1,1]$, hence triangular:
\begin{equation}
  f(u) = \frac{2-|u|}{4}\ \ (|u|\le2), \qquad f(0)=\tfrac12 .
\end{equation}
$J$ is uniform on $[-1,1]$ and independent of both.

Since $R=O(t^2)$ is small, the condition \eqref{eq:app:res} forces
$v\simeq\mp u$, and on that locus the two terms in $R$ coincide, giving
\begin{equation}
  R\big|_{v=-u} \;=\; \frac{t^2|J|}{|u^2-J^2|} .
\end{equation}
Now, $R<t$, so the probability that something is $<R$ can be approximated with its density at the origin times $R$, namely $p(0)R$. At $u\simeq \pm v$, $p(0)=f(u)^2$ and $f(u)=(2-|u|)/4$.

Evaluated in this way, the probability of one of the eight conditions is
\begin{equation}
  P_1 \;=\; 2\int_{-1}^{1}\!\frac{dJ}{2}\int_B\! du\, f(u)^2\,
            \frac{t^2|J|}{|u^2-J^2|}
      \;=\; t^2\int_{-1}^{1}\! dJ\,|J|\,\mathcal{I}(J),
  \qquad
  \mathcal{I}(J)=\int\! du\,\frac{f(u)^2}{|u^2-J^2|} .
\end{equation}
where $B=\{u\in[-2,2]\,:\,|u+J|>t/2,|u-J|>t/2\}$ because the denominators have been filtered for failure at step 1 and are therefore larger than $t/2$.   

The integrals can be done explicitly and then expanded for small $t$. A $\log(t)$ term is generated
by the ``cutoff" of $t/2$ due to the near-resonances that were just survived at the previous scale.

Writing $a=|J|$, $\varphi(u)=f(u)^2=(2-|u|)^2/16$ and
$\psi(u)=\varphi(u)/(u+a)$, and using the symmetry $u\to-u$,
\begin{equation}
  a\,\mathcal{I}(a) \;=\; \varphi(a)\,\ln\!\frac{4a(2-a)}{t^2}
                       \;+\; 2a\,\mathrm{Reg}(a),
  \qquad
  \mathrm{Reg}(a)=\int_0^2\!\frac{\psi(u)-\psi(a)}{|u-a|}\,du ,
\end{equation}
the remainder being finite as the cutoff is removed. Hence (to $O(t^2)$)
\begin{equation}
  P_1 = t^2\Big[\,4\ln(1/t)\!\int_0^1\!\varphi\,da
        \;+\; 2\!\int_0^1\!\varphi(a)\ln\!\big(4a(2-a)\big)da
        \;+\; 2\!\int_0^1\! 2a\,\mathrm{Reg}(a)\,da \Big].
\end{equation}
The first integral is elementary,
$\int_0^1\varphi\,da=\tfrac1{16}\int_0^1(2-a)^2da=\tfrac{7}{48}$, and the other
two are evaluated
\begin{eqnarray}
  \int_0^1\!\varphi(a)\ln\big(4a(2-a)\big)\,da &=&\frac{1}{144} (66 \log (2)-35)=0.07464,\nonumber\\
  \int_0^1\! 2a\,\mathrm{Reg}(a)\,da &=&\frac{1}{48} (-7+46 \log (2)-27 \log (3))=-0.09954 .
\end{eqnarray}
So
\begin{equation}
    P_1=\frac{7}{12}t^2\ln(c_2/t)+o(t^2),
\end{equation}
with $c_2=0.918...$\ .

Considering there were 8 independent conditions and a factor of $1/2$ from the BCH formula which reduces $P_1$ by $1/2$ (assuming $t\ll 1$, the probability is linear in the amplitude), the rate $a_2$ is $a_2=4P_1$. This is valid {\it assuming $P_1\ll 1$} and that we can use a rare events, Poisson statistics hypothesis. The correction would be $O(P_1^2)=O(t^4)$ and therefore negligible.

The final result, in this approximation, is
\begin{equation}
    a_2(t)=\frac{7}{3}t^2\ln(c_2/t).
    \label{eq:app:a2FWSA}
\end{equation}
Numerically $a_2(t)\propto t^{2.00\pm0.06}$ (Fig.~\ref{fig:ak_t}), the deviation from $t^2$ is smaller than what the logarithm in \eqref{eq:app:a2FWSA} suggests.

For higher order $a_k$'s the small $t$ approximation gives the results
$\propto t^k\ln(c_k/t)$. This is apparently in contradiction with an expected
Newton--KAM $t^{2^k}$ correction. The reason for this is that while the
typical values of the numerators decay, the typical value of the denominators also decay, and iterating the typical value of $V_k$ to be eliminated (which has a long-tail Pareto distribution) powers of $t$ almost compensate. To give a schematic, order of magnitude explanation of this phenomenon let us make up a toy random process model.

\subsection{A toy model of the flow}
\label{app:toy}

In this subsection we replace the operator recursion by a recursion for a few
scalar random variables. Each one stands for the order of magnitude of a
coefficient of one operator of the flow. We first recall the exact objects,
then give the dictionary, and finally list the approximations.

\paragraph{Exact objects.}
At scale $k$ the Hamiltonian is $H^{(k)}=E^{(k)}+V^{(k)}$, with
$E^{(0)}=E=\sum_i h_iZ_i+\sum_i J_iZ_iZ_{i+1}$ and $V^{(0)}=t\sum_iX_i$.
Any off-diagonal operator is expanded in $X$-monomials,
\begin{equation}
V^{(k)}=\sum_{S\neq\emptyset}X_S\,v^{(k)}_S(Z),\qquad X_S=\prod_{i\in S}X_i ,
\end{equation}
where $v^{(k)}_S(Z)$ is a function of the $Z$'s, i.e.\ a diagonal operator.
The denominator of the monomial $S$ is the diagonal operator
$\partial_SE\equiv X_SEX_S-E$. For a contiguous $S=\{a,\dots,b\}$ the
interior bonds are invariant under $X_S$, and
\begin{equation}
\partial_SE=-2\Big(\sum_{i=a}^{b}h_iZ_i+J_{a-1}Z_{a-1}Z_a+J_bZ_bZ_{b+1}\Big).
\label{eq:dSE}
\end{equation}
It depends on the $|S|+2$ spins $Z_{a-1},\dots,Z_{b+1}$. Under
$\sigma\to X_S\sigma$, i.e.\ exchanging the two states connected by $X_S$,
it only changes sign. Hence $|\partial_SE|$ depends on the $|S|$ internal
spins up to a global flip and on the two boundary spins, and takes
$2^{|S|+1}$ distinct values: one per pair of states $(\sigma,X_S\sigma)$
connected by $X_S$.

The generator of scale $k$ solves $[A^{(k)},E^{(k-1)}]=-V^{(k-1)}$, i.e.
\begin{equation}
A^{(k)}=\sum_S X_S\,a^{(k)}_S(Z),\qquad
a^{(k)}_S(\sigma)=\frac{v^{(k-1)}_S(\sigma)}{\partial_SE^{(k-1)}(\sigma)} .
\end{equation}
The angle of scale $k$ is
$\theta_k=\max_{S,\sigma}|a^{(k)}_S(\sigma)|$, and scale $k$ is resonant if
$\theta_k>1$. Using $[A^{(k)},E^{(k-1)}]=-V^{(k-1)}$ in the BCH expansion gives
\begin{equation}
V^{(k)}=\sum_{n\ge1}\frac{n}{(n+1)!}\,
\mathrm{ad}^{\,n}_{A^{(k)}}V^{(k-1)}
=\tfrac12\,[A^{(k)},V^{(k-1)}]+O\big(A^2V\big),
\label{eq:BCHtoy}
\end{equation}
where the diagonal part ($S=\emptyset$, generated by $X_SX_S=\mathbb{I}$) is
moved into $E^{(k)}$.

\paragraph{Dictionary.}
Fix one monomial $S$ and one configuration $\sigma$ that are dangerous at
scale $k$. The toy variables are
\begin{center}
\begin{tabular}{lll}
\hline
toy variable & operator & meaning \\
\hline
$y_k$ & $|v^{(k-1)}_S(\sigma)|$ & numerator: coefficient of $V^{(k-1)}$; $y_1=t$ \\
$z_k$ & $|\partial_SE^{(k-1)}(\sigma)|$ & denominator \\
$\theta_k=y_k/z_k$ & $|a^{(k)}_S(\sigma)|$ & coefficient of $A^{(k)}$: the angle \\
$y_{k+1}=\tfrac12\theta_k y_k$ & $\tfrac12[A^{(k)},V^{(k-1)}]$ & numerator of the next scale \\
$N_k$ & $\#\{(S,\sigma)\}$ per site & number of competing denominators \\
$m_k$ & & number of dangerous monomials of scale $k+1$ fed by one of scale $k$ \\
$\Lambda$ & $1/\rho(0)$ & $\rho$: density of $|\partial_SE|$ at $0$ \\
\hline
\end{tabular}
\end{center}
Along one chain of dangerous monomials (a {\it history}) the toy recursion is
\begin{equation}
y_{k+1}=\frac12\,\frac{y_k^2}{z_k},\qquad y_1=t .
\label{eq:toyrec}
\end{equation}

\paragraph{Approximations.}
\begin{itemize}
\item[(T1)] \emph{Range equals scale.} A coefficient of order $t^k$ contains
$k$ factors $X_i$ from $V^{(0)}$, so its support has at most $k$ sites. We
take the dangerous monomial of scale $k$ to have $|S|=k$. By
Eq.~\eqref{eq:dSE} this gives $N_k=2^{k+1}$, so $N_{k+1}/N_k=2$.
\item[(T2)] \emph{BCH truncated at $n=1$} in Eq.~\eqref{eq:BCHtoy}.
\item[(T3)] \emph{Magnitudes multiply, multiplicities are counted
phenomenologically.} Along a history, the coefficient of $[A,V]$ is the
product of the coefficients of $A$ and $V$. A dangerous monomial of scale $k$,
however, does not feed a single monomial of scale $k+1$: in the commutator
$[A^{(k)},V^{(k-1)}]$ it is combined with the lower-scale monomials it overlaps,
and an order-$t^k$ coefficient collects the different orderings of the flips that
build it. We assume that each dangerous monomial of scale $k\geq2$ feeds
\begin{equation}
m_k=c\,(k-1)
\label{eq:mk}
\end{equation}
dangerous monomials of scale $k+1$, and that $m_1=c$ as well, so that every step carries the same factor $c$. (The leading-order calculation of $a_2$ above corresponds to $m_1=1$; the measured $a_2$ is indeed a factor $2$--$2.5$ larger than \eqref{eq:app:a2FWSA} at $t=0.025$--$0.045$.) The factor $k-1$ is the number of lower scales whose monomials
can be combined with it, and $c$ is a phenomenological constant that lumps all the
remaining multiplicities and cancellations. The number of histories that reach
scale $k$ from one site is then
\begin{equation}
M_k=\prod_{j=1}^{k-1}m_j=c^{\,k-1}\,(k-2)!,\qquad k\geq2 .
\label{eq:Mk}
\end{equation}
Taking $m_k=1$ instead (a single history per site) removes the multiplicity
altogether; we comment on this case at the end.
This is what we observe in the test of the $\epsilon$ dependence in Sec.~\ref{sec:model}: lowering the cutoff from $10^{-6}$ to $10^{-7}$ raises the fraction of samples resonating at step $3$ from $14.4\%$ to $17.7\%$, mostly through samples that converge at the larger cutoff, because contributions discarded at earlier scales, each below $\epsilon$, add up on the same strings at later ones. A dangerous monomial of scale $k+1$ is thus fed by monomials of all the previous scales, not by a single chain, which is what the factor $k-1$ in \eqref{eq:mk} accounts for. The same lower-scale contributions also renormalize the energies and can lower the denominators, an effect neglected here by (T4).
\item[(T4)] \emph{Unrenormalised denominators}: $E^{(k-1)}\to E$.
\item[(T5)] \emph{Uniform independent denominators}: the $N_k$ values of
$z_k$ are i.i.d.\ uniform on $[0,\Lambda]$ and independent of the previous
scales and of the other histories. Only the density at the origin matters,
because a resonance needs $z_k<y_k\ll1$.
\end{itemize}

\paragraph{Hazard.}
A history is resonant at scale $k$ if the smallest of its $N_k$ denominators is
below $y_k$:
\begin{equation}
q_k=1-\Big(1-\frac{y_k}{\Lambda}\Big)^{N_k}\simeq1-e^{-u_k}\simeq u_k(1+O(t)),\qquad
u_k\equiv\frac{N_k\,y_k}{\Lambda}\ll 1.
\end{equation}
$u_k$ is the numerator measured in units of the typical smallest denominator,
$\Lambda/N_k$. If scale $k$ is not resonant, the relevant denominator is the
smallest one, $z_k=\min_j z_k^{(j)}>y_k$, since it gives the largest angle
$\theta_k$ and the largest $y_{k+1}$. We write it as
\begin{equation}
z_k=y_k+\frac{\Lambda}{N_k}\,\varepsilon_k ,
\end{equation}
so $\varepsilon_k$ is the gap between the smallest denominator and the
numerator, in units of the typical spacing $\Lambda/N_k$. Conditioned on
$z_k>y_k$, $\mathbb P(\varepsilon_k>x)=\big(1-x/[N_k(1-y_k/\Lambda)]\big)^{N_k}
\to e^{-x}$ for $N_k\gg1$, i.e.\ $P(\varepsilon_k)=e^{-\varepsilon_k}$. Only its
density at the origin, which equals $1$ for every $N_k$, enters the coefficient of the logarithm.

Inserting this in \eqref{eq:toyrec},
\begin{equation}
u_{k+1}=\frac{N_{k+1}}{N_k}\cdot\frac12\cdot\frac{u_k^2}{u_k+\varepsilon_k}
=\frac{u_k^2}{u_k+\varepsilon_k}.
\label{eq:urec}
\end{equation}
The factor $\tfrac12$ of BCH is cancelled exactly by the doubling of the
number of denominators, and along a history the only parameter left is $u_1$.
Since $u_{k+1}<u_k$, every history satisfies $u_k\le u_1=\tfrac32t\ll1$, so
$q_k=u_k[1+O(t)]$ uniformly. The histories of one site are independent (T5) and
rare, and a chain of length $L$ has $L$ independent starting sites, so
$S_k=e^{-a_kL}$ with
\begin{equation}
a_k\simeq M_k\,\langle u_k\rangle ,
\end{equation}
where $\langle\cdot\rangle$ is the average over the non-resonant history
$\varepsilon_1,\dots,\varepsilon_{k-1}$ of a single history. The average is not
dominated by the typical value, which is doubly exponentially small, but by rare,
nearly resonant histories.

The relation $a_k\simeq M_k\langle u_k\rangle$ is the same as that between the resonance probability of a single path and that of a site at distance $r$ in the Anderson problem \cite{Anderson58}: exponentially many paths lead to that site and, treating them as independent, one takes the maximum over them, or their sum, which is dominated by the maximum. On the Bethe lattice \cite{abou1973selfconsistent} the exponential number of paths is replaced by the exponential number of sites at distance $r$. The same counting underlies the forward scattering approximation \cite{pietracaprina2016forward}, the construction of local integrals of motion in \cite{ros2015integrals} and, originally, the perturbative analysis of BAA \cite{basko2006metal}; here the histories play the role of the paths and $M_k$ that of their number.

\paragraph{Calibration at $k=1$.}
For $S=\{i\}$, $|\partial_iE|=2|h_i\pm J_{i-1}\pm J_i|$. The sum of three
variables uniformly distributed in $[-1,1]$ has density $3/8$ at the origin, so $\rho(0)=3/8$ and $\Lambda=8/3$. With $N_1=4$,
\begin{equation}
u_1=\frac{N_1t}{\Lambda}=\frac32\,t=q_1 +O(t^2),
\end{equation}
which reproduces $a_1=\tfrac32t$ exactly.

\paragraph{Moments of a single history.}
From \eqref{eq:urec} and $P(\varepsilon)=e^{-\varepsilon}$,
\begin{equation}
    \langle u_{k+1}^s\,|\,u_k\rangle=u_k^{2s}\int_0^\infty d\varepsilon\, \frac{e^{-\varepsilon}}{(u_k+\varepsilon)^s}.
\end{equation}
For $s>1$ and $u_k\ll 1$ the integral is dominated by $\varepsilon\lesssim u_k$ and equals $u_k^{1-s}/(s-1)$, so
\begin{equation}
    \langle u_{k+1}^s\rangle=\frac{\langle u_k^{s+1}\rangle}{s-1},
\end{equation}
which, iterated from the deterministic $u_1$, gives
\begin{equation}
    \langle u_{k}^s\rangle=u_1^{\,s+k-1}\,\frac{\Gamma(s-1)}{\Gamma(s+k-2)} .
\end{equation}
For $s=1$ the integral is $e^{u_k}E_1(u_k)\simeq\ln(1/u_k)-\gamma$, with $\gamma$ the Euler--Mascheroni constant, so
\begin{equation}
    \langle u_{k+1}\rangle=\big\langle u_k^2\ln (1/u_k)\big\rangle-\gamma\langle u_k^2\rangle .
\end{equation}
Using $\langle u_k^2\ln (1/u_k)\rangle=-\partial_s\langle u_k^s\rangle|_{s=2}=\langle u_k^2\rangle\,\big(\ln(1/u_1)+H_{k-1}\big)$, where $H_n=\sum_{m=1}^{n}1/m$ is the harmonic number ($H_0=0$), and shifting $k\to k-1$,
\begin{equation}
\langle u_k\rangle=\frac{u_1^{\,k}}{(k-2)!}
\left[\ln\frac{1}{u_1}+H_{k-2}-\gamma\right],\qquad k\ge2 .
\label{eq:uk_mean}
\end{equation}
The factor $1/(k-2)!$ is the probability that a single history goes through a nested sequence of near-resonances, each closer than the previous one.

\paragraph{Rates.}
Multiplying by the number of histories \eqref{eq:Mk}, the factorial cancels and, with $u_1=\tfrac32t$,
\begin{equation}
    a_k=c^{\,k-1}\Big(\frac{3}{2}t\Big)^{k}\left[\ln\frac{2}{3t}+H_{k-2}-\gamma\right],\qquad k\geq2 .
\label{eq:ak_toy}
\end{equation}
For $k=2$ and $c=1$, \eqref{eq:ak_toy} is already a good approximation of the leading-order result, with the coefficient $7/3$ of the logarithm replaced by $(3/2)^2=9/4$. For larger $k$ the rates behave as $a_k\sim t^k\ln(1/t)$, the multiplicity entering only through the prefactor $c^{k-1}$.

\paragraph{Transition.}
The ratio of the first two rates is $a_2/a_1=c\,u_1\big[\ln(1/u_1)-\gamma\big]$, and for $k\geq2$
\begin{equation}
    \frac{a_{k+1}}{a_k}=\frac32\,c\,t\;\frac{\ln\frac{2}{3t}+H_{k-1}-\gamma}{\ln\frac{2}{3t}+H_{k-2}-\gamma}\;\xrightarrow[k\to\infty]{}\;\frac32\,c\,t ,
\end{equation}
since $H_{k-1}-H_{k-2}=1/(k-1)\to0$. The series $a(t)=\sum_k a_k(t)$ is therefore geometric at large $k$: it converges for $t<t_c$ and diverges for $t>t_c$, with
\begin{equation}
    t_c=\frac{2}{3c}.
\end{equation}
At $t_c$ the rates become scale invariant, $a_{k+1}=a_k$ up to the slow logarithmic factor, which is the avalanche condition discussed in Sec.~\ref{sec:numerics}: resonances appear with the same probability at every scale, and $P_{\rm conv}$ decays faster than any exponential in $L$. In this toy model the transition is a competition between the factor $\tfrac32t$ lost at each scale by the numerators and the growing number $c\,(k-1)$ of ways in which a dangerous monomial can feed the next scale. The derivation assumes $u_1=\tfrac32t\ll1$; at $t_c$ this requires $c\gg1$, in which case the rates at the transition, $a_k(t_c)=c^{-1}\big[\ln c+H_{k-2}-\gamma\big]$, are also small, consistently with the Poisson counting.

\begin{figure}[!htbp]
    \centering
    \includegraphics[width=0.75\linewidth]{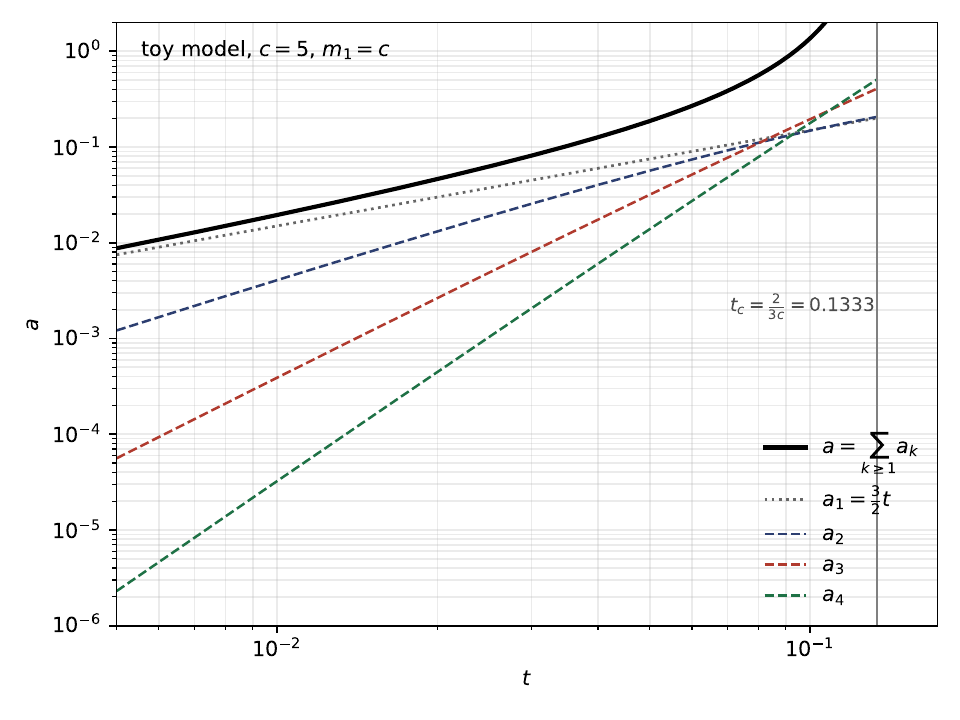}
    \caption{Total rate $a=\sum_{k\geq1}a_k$ (solid) of the toy model with multiplicity $c=5$ ($m_k=c(k-1)$ for $k\geq2$, $m_1=c$), with $a_1=\tfrac32t$ (dotted) and $a_2$, $a_3$, $a_4$ from \eqref{eq:ak_toy} (dashed). The series diverges at $t_c=2/(3c)$ (vertical line).}
    \label{fig:toy}
\end{figure}

Fig.~\ref{fig:toy} shows the total rate $a(t)=\sum_{k\geq1}a_k(t)$ and the first four rates for $c=5$, for which $t_c=2/15\simeq0.133$. The sum can be done in closed form using $\sum_{n\geq0}x^n=1/(1-x)$ and $\sum_{n\geq0}H_nx^n=-\ln(1-x)/(1-x)$ with $x=cu_1=t/t_c$:
\begin{equation}
    a(t)=\frac32\,t+\frac{\tfrac94\,c\,t^2\Big[\ln\frac{2}{3t}-\gamma-\ln\big(1-\frac{t}{t_c}\big)\Big]}{1-t/t_c},
\label{eq:atot_toy}
\end{equation}
which diverges as $\ln\!\big[1/(1-t/t_c)\big]/(1-t/t_c)$ at $t_c$. The crossings $a_2=a_3$, $a_1=a_3$ and $a_1=a_2$ occur at $t\simeq0.081$, $0.084$ and $0.104$, in the same order as the measured ones (Fig.~\ref{fig:ak_t}) and, for this value of $c$, at nearby values of $t$; the crossings of the first few rates precede the divergence of the series. With this $c$ the toy model overestimates the measured $a_2$ and $a_3$ by factors of about $1.4$ and $2$, the values $c\simeq3$--$3.7$ that reproduce their magnitude giving no crossing of $a_1$ and $a_2$ below $t\simeq0.1$.

Without the multiplicity ($m_k=1$) the rates are $a_k=\langle u_k\rangle$ of \eqref{eq:uk_mean}, the ratio $a_{k+1}/a_k\simeq\tfrac32t/(k-1)$ vanishes at large $k$, and $a(t)$ converges for every $t$: there is no transition. The same is true if the multiplicity is a constant, $m_k=c$, since the factorial is then not compensated. A transition at finite $t$ requires that the number of ways a dangerous monomial can feed the next scale grows with the scale. This is also behind the difficulty encountered with the $\epsilon$ cutoff. Any cutoff can remove a string which at step $k$ has very little weight but at step $k+n$ could see its weight increase manyfold because of a rare resonance. Controlling such rare events is the main difficulty that any analytic proof must tackle.

Below $t_c$ the rates are not monotonic in $k$. Writing $x=t/t_c$ and $B_k=\ln\frac{2}{3t}+H_{k-2}-\gamma$, \eqref{eq:ak_toy} gives $a_{k+1}/a_k=x\,[1+1/((k-1)B_k)]$: the rates grow as long as $(k-1)(1-x)B_k<x$ and decrease afterwards, so that they have a maximum at a scale $k^*(t)$. Far from the transition the maximum is at the first scales (for $c=5$, $a_1$ is the largest rate for $t\lesssim0.6\,t_c$); approaching $t_c$ it moves to larger $k$, with $k^*\sim1/\big[(1-t/t_c)\ln\frac{1}{1-t/t_c}\big]$ diverging at $t_c$. Resonances are thus concentrated on the scales up to $k^*$, beyond which the rates decay geometrically with ratio $t/t_c$; at $t_c$ this scale becomes infinite and the rates are scale invariant. The avalanche process at the transition continues forever.

\section{Conclusions and Discussion}

\subsection{What did we do?}

We begin by recapitulating the main technical results of this paper.
By implementing the Schrieffer--Wolff flow in DRGHP \cite{DRGHP} to system sizes $L\leq 32$, we have 
checked that the probability of convergence is only exponentially small in $L$ and estimated a rate per spin $a(t)$ and separate rates $a_k(t)$ of failure at step $k$ (but not before). The numerical estimates give two important results: first, they show that the DRGHP bound is loose by a factor of 10-15 on the per spin total rate $a(t)$, which translates into many orders of magnitude on the probability itself; Second, the separate analysis of the rates $a_k(t)$ points at an avalanche transition at $t_c\simeq 0.08$. The code was developed and tested, and the small sizes and the cross-checks were run, on a single 8-core Intel Xeon workstation with 256GB of RAM. The production runs at $L=16$--$32$ were done on the Galileo100 cluster at CINECA, for a total of about $3.6\times10^4$ core hours. Our numerics should still be considered a proof of principle rather than a large scale, definitive numerical work. In fact, a natural continuation of this work would consist of finding, accurately, $a_k(t)$ for $k\geq 4$ (at least to $k=6$), which is doable with our code and a larger allocation of computer time. For colleagues interested in pursuing this, all of our code, written in C++ with considerable help from Claude AI, is available upon request, both to enable substantially larger runs and to facilitate independent scrutiny and replication of the calculation. We would particularly welcome independent checks of the implementation and, where useful, independent reimplementations of the calculation. 

\subsection{Why did we do it?}

As we noted in the Introduction, we returned to the Griffiths-MBL problem following DRGHP because the field had failed to reach consensus on the full MBL problem. We feel it may be useful to readers to sketch in slightly more detail the genesis of this work.

This story begins in 2019, when Šuntajs, Bonča, Prosen, and Vidmar presented numerical work challenging the existence of many-body localization even in one dimension, and arguing that the previous consensus, believed to be supported by other numerical work and approximate arguments, was incorrect. This was contested by others in the field, who noted that workers had been quite aware of the challenges with the numerical evidence, but they also acknowledged that what the problem lacked was convincing evidence, at large but finite disorder values, that the phenomenon was stable in the infinite-volume limit. Subsequent work by Sels and Polkovnikov sharpened this challenge from a somewhat different direction, further contributing to the motivation for revisiting the problem.

As happens in those not-so-common cases in which the community of working physicists, armed with various non-rigorous techniques, cannot agree on the answer to a question, it becomes necessary to bring greater mathematical firepower to bear and look for a rigorous proof. The somewhat odd situation in 2019 was that John Imbrie had already published such a proof three years earlier, conditional on an assumption concerning level attraction.\footnote{Posted on arXiv five years earlier.}

So why was this proof not sufficient for the purposes of the community? There were two related reasons. The first was that when the proof appeared, most of the evidence seemed aligned, and so it carried considerable authority and there was relatively little incentive for an independent adversarial reconstruction of its details. It was a long and complicated proof, appropriate to a problem with an infinite hierarchy of energy scales. Our impression, based also on discussions with experts in the area, was that while the proof was taken very seriously, confidence in it should not be confused with the existence of multiple independent line-by-line checks of the argument. The second difficulty was that the techniques involved were sufficiently far from those familiar to most working many-body physicists that the proof was difficult for the broader community to assimilate independently.

Given the situation in 2019, a group of us gathered at the GGI in Florence, at an INFN meeting organized by one of us (AS) and asked John Imbrie, who was present, if he would try to explain his proof to us non-mathematical physicists. He made a valiant attempt, but unfortunately his audience, in which one of us (SLS) was present (AS being absent on account of some health issues), was starting too far from the techniques he was using to be able to convince itself that the proof had already settled the question. This exercise led one of us to the conclusion that what was needed was a determined effort by a subset of the community to unpack the proof and try to check parts of it using techniques with which we were more familiar. Unfortunately, COVID intervened, and the idea was put on hold.

The idea returned when DRGHP announced their proof of the Griffiths-MBL problem, and SLS, together with Vedika Khemani, convened a small specialist meeting at Oxford in August 2024. During the bulk of a week, De Roeck and Huveneers explained at some length their main ideas and patiently responded to questions from the audience. As this addressed a significantly more bounded problem, with one central probability estimate for nonresonant regions of a given size, it became clear that one way for us to engage with it would be to calculate this probability directly and compare it with the bounds used in the paper. As the reader has seen, that is what we have done here. We believe this increases confidence in their proof and another independent calculation to even larger system sizes would improve confidence even further.

More broadly, the next step is clearly to engage in a similar fashion either with the original mathematical-physics proof, or with a forthcoming proof by De Roeck, Huveneers, and collaborators, which will provide another route to the same result. While this will be a steeper hill to climb, advances in machine intelligence—which we benefited from enormously in the present work—should make it substantially easier for the physics community to engage in such checks of the key parts of these proofs.

Now, our discussion in the last few paragraphs is not one that one would traditionally see in a physics journal, although it is perfectly possible to encounter it in discussions of the philosophy and sociology of physics. But physics, like science in general, is done by fallible workers, and confidence in its results has always depended upon experimental or theoretical replication, ideally using independent techniques. What we have done here is simply to make that process of Bayesian updating explicit, and to carry out part of it.

\subsection{A strategy for the era of machine generated proofs}

We would like to end with the observation that approaching announced major results in this deliberate fashion is likely something that the theoretical-physics community will have to adopt more broadly as a mode of operation. Extraordinarily powerful AI models have been built as we write this, and the age of machine-constructed proofs of significant results appears to be dawning.

At least to begin with, machine proofs seem most likely to be novel on problems where progress requires combining a large number of techniques and mastering a great deal of detail. It is also likely that such proofs will themselves be machine-formalized and checked by proof assistants. We could therefore end up with machine-generated proofs whose formalizations are themselves formally verified, but whose content most physicists will nevertheless find difficult to assimilate. Such checking can provide exceptionally strong assurance that a formal theorem follows from its stated formal assumptions, but it does not by itself guarantee that an automatically generated formalization faithfully captures the original informal statement and argument, still less that it supplies physical understanding of why the result is true. Terence Tao has commented on this possibility from the perspective of mathematics where, {\it mutatis mutandis}, the same challenge is likely to arise.

As this starts to happen (while we were completing this work, \cite{OpenAINS2026} happened) it seems to us that the theoretical-physics community will need to engage with such proofs much as we have engaged, in microcosm, with the DRGHP proof here: by independently interrogating both the correspondence between the formal statement and the physical claim and a sufficient number of the key steps to develop a much better appreciation of what the proof is actually doing. This interrogation need not be purely numerical. It can also proceed analytically, by isolating manageable pieces of a long argument and rederiving or checking them with more familiar methods, potentially with the assistance of AI systems themselves. Numerical implementations, independent code audits, analytic checks of selected steps, and, where useful, independent reimplementations can all contribute to this process. Ideally, such engagement will generate fruitful ideas for further work, or perhaps enable the more mathematically minded to construct simpler and more human-assimilable proofs.

\section{Acknowledgements}

We thank Wojciech De Roeck and Fran\c{c}ois Huveneers for the considerable time and effort they devoted to explaining the ideas and technical structure of the DRGHP proof at the Oxford meeting in August 2024 and for feedback and questions on the first version of this draft, and Vedika Khemani for helping to convene the 2024 meeting. We also thank the other participants for the enthusiasm, curiosity, and persistence with which they engaged in the collective effort to understand and interrogate the proof.

The work of AS has also been supported by the European Union--NextGenerationEU under the NRRP Project ``National Quantum Science and Technology Institute'' (NQSTI), Award Number PE00000023, Concession Decree No.~1564 of 11.10.2022 adopted by the Italian Ministry of Research, CUP J97G22000390007.

We acknowledge a CINECA award under the ISCRA initiative, for the availability of high performance computing resources and support (project IscrC\_QuDyDi).

\bibliography{references}

@article{DRGHP,
  title={Absence of normal heat conduction in strongly disordered interacting quantum chains},
  author={De Roeck, Wojciech and Giacomin, Lydia and Huveneers, Francois and Prosniak, Oskar},
  journal={arXiv preprint arXiv:2408.04338},
  year={2024}
}

@article{Anderson58,
  author = {Anderson, P W},
  title = {Absence of diffusion in certain random lattices},
  journal = {Phys. Rev.},
  volume = {109},
  pages = {1492--1505},
  year = {1958},
  doi = {10.1103/PhysRev.109.1492},
}

@article{basko2006metal,
  title={Metal--insulator transition in a weakly interacting many-electron system with localized single-particle states},
  author={Basko, Denis M and Aleiner, Igor L and Altshuler, Boris L},
  journal={Annals of physics},
  volume={321},
  number={5},
  pages={1126--1205},
  year={2006},
  publisher={Elsevier}
}

@article{nandkishore2015many,
  title={Many-body localization and thermalization in quantum statistical mechanics},
  author={Nandkishore, Rahul and Huse, David A},
  journal={Annu. Rev. Condens. Matter Phys.},
  volume={6},
  number={1},
  pages={15--38},
  year={2015},
  publisher={Annual Reviews}
}

@article{sierant2025many,
  title={Many-body localization in the age of classical computing},
  author={Sierant, Piotr and Lewenstein, Maciej and Scardicchio, Antonello and Vidmar, Lev and Zakrzewski, Jakub},
  journal={Reports on Progress in Physics},
  volume={88},
  number={2},
  pages={026502},
  year={2025},
  publisher={IOP Publishing}
}

@article{Imbrie16,
  author = {Imbrie, John Z},
  title = {Diagonalization and many-body localization for a disordered quantum spin chain},
  journal = {Phys. Rev. Lett.},
  volume = {117},
  pages = {027201},
  year = {2016},
  doi = {10.1103/PhysRevLett.117.027201},
}

@article{Imbrie16a,
  author = {Imbrie, John Z},
  title = {On many-body localization for quantum spin chains},
  journal = {Journal of Statistical Physics},
  volume = {163},
  pages = {998--1048},
  year = {2016},
  doi = {10.1007/s10955-016-1508-x},
}

@article{Pal10,
  author = {Pal, Arijeet and David A.\ Huse},
  title = {Many-body localization phase transition},
  journal = {Phys. Rev. B},
  volume = {82},
  pages = {174411},
  year = {2010},
  doi = {10.1103/PhysRevB.82.174411},
}

@article{smith2016many,
  title={Many-body localization in a quantum simulator with programmable random disorder},
  author={Smith, Jacob and Lee, Aaron and Richerme, Philip and Neyenhuis, Brian and Hess, Paul W and Hauke, Philipp and Heyl, Markus and Huse, David A and Monroe, Christopher},
  journal={Nature Physics},
  volume={12},
  number={10},
  pages={907--911},
  year={2016},
  publisher={Nature Publishing Group UK London}
}

@article{bardarson2012unbounded,
  title={Unbounded growth of entanglement in models of many-body localization},
  author={Bardarson, Jens H and Pollmann, Frank and Moore, Joel E},
  journal={Physical review letters},
  volume={109},
  number={1},
  pages={017202},
  year={2012},
  publisher={APS}
}

@article{ros2015integrals,
title = {Integrals of motion in the many-body localized phase},
journal = {Nuclear Physics B},
volume = {891},
pages = {420-465},
year = {2015},
issn = {0550-3213},
doi = {https://doi.org/10.1016/j.nuclphysb.2014.12.014},
url = {https://www.sciencedirect.com/science/article/pii/S0550321314003836},
author = {V. Ros and M. Müller and A. Scardicchio}
}

@article{iyer2013many,
  title={Many-body localization in a quasiperiodic system},
  author={Iyer, Shankar and Oganesyan, Vadim and Refael, Gil and Huse, David A},
  journal={Physical Review B—Condensed Matter and Materials Physics},
  volume={87},
  number={13},
  pages={134202},
  year={2013},
  publisher={APS}
}

@article{alet2018many,
  title={Many-body localization: An introduction and selected topics},
  author={Alet, Fabien and Laflorencie, Nicolas},
  journal={Comptes Rendus. Physique},
  volume={19},
  number={6},
  pages={498--525},
  year={2018}
}

@article{imbrie2017local,
  title={Local integrals of motion in many-body localized systems},
  author={Imbrie, John Z and Ros, Valentina and Scardicchio, Antonello},
  journal={Annalen der Physik},
  volume={529},
  number={7},
  pages={1600278},
  year={2017},
  publisher={Wiley Online Library}
}

@article{schulz2020phenomenology,
  title={Phenomenology of anomalous transport in disordered one-dimensional systems},
  author={Schulz, Maximilian and Taylor, Scott R and Scardicchio, Antonello and {\v{Z}}nidari{\v{c}}, Marko},
  journal={Journal of Statistical Mechanics: Theory and Experiment},
  volume={2020},
  number={2},
  pages={023107},
  year={2020},
  publisher={IOP Publishing and SISSA}
}

@article{panda2019can,
  title={Can we study the many-body localisation transition?},
  author={Panda, Rajat K and Scardicchio, Antonello and Schulz, Maximilian and Taylor, Scott R and {\v{Z}}nidari{\v{c}}, Marko},
  journal={Europhysics Letters},
  volume={128},
  number={6},
  pages={67003},
  year={2019},
  publisher={EDP Sciences, IOP Publishing and Societ{\`a} Italiana di Fisica}
}

@article{agarwal2017rare,
  title={Rare-region effects and dynamics near the many-body localization transition},
  author={Agarwal, Kartiek and Altman, Ehud and Demler, Eugene and Gopalakrishnan, Sarang and Huse, David A and Knap, Michael},
  journal={Annalen der Physik},
  volume={529},
  number={7},
  pages={1600326},
  year={2017},
  publisher={Wiley Online Library}
}

@article{Agarwal2015,
  author  = {Kartiek Agarwal and Sarang Gopalakrishnan and Michael Knap and Markus M{\"u}ller and Eugene Demler},
  title   = {Anomalous Diffusion and Griffiths Effects Near the Many-Body Localization Transition},
  journal = {Physical Review Letters},
  volume  = {114},
  number  = {16},
  pages   = {160401},
  year    = {2015},
  doi     = {10.1103/PhysRevLett.114.160401}
}

@article{vznidarivc2016diffusive,
  title={Diffusive and subdiffusive spin transport in the ergodic phase of a many-body localizable system},
  author={{\v{Z}}nidari{\v{c}}, Marko and Scardicchio, Antonello and Varma, Vipin Kerala},
  journal={Physical review letters},
  volume={117},
  number={4},
  pages={040601},
  year={2016},
  publisher={APS}
}

@article{oganesyan2007localization,
  title={Localization of interacting fermions at high temperature},
  author={Oganesyan, Vadim and Huse, David A},
  journal={Physical Review B—Condensed Matter and Materials Physics},
  volume={75},
  number={15},
  pages={155111},
  year={2007},
  publisher={APS}
}

@article{vsuntajs2020quantum,
  title={Quantum chaos challenges many-body localization},
  author={{\v{S}}untajs, Jan and Bon{\v{c}}a, Janez and Prosen, Toma{\v{z}} and Vidmar, Lev},
  journal={Physical Review E},
  volume={102},
  number={6},
  pages={062144},
  year={2020},
  publisher={APS}
}

@article{morningstar2022avalanches,
  title={Avalanches and many-body resonances in many-body localized systems},
  author={Morningstar, Alan and Colmenarez, Luis and Khemani, Vedika and Luitz, David J and Huse, David A},
  journal={Physical Review B},
  volume={105},
  number={17},
  pages={174205},
  year={2022},
  publisher={APS}
}

@article{pietracaprina2016forward,
  title={Forward approximation as a mean-field approximation for the Anderson and many-body localization transitions},
  author={Pietracaprina, Francesca and Ros, Valentina and Scardicchio, Antonello},
  journal={Physical Review B},
  volume={93},
  number={5},
  pages={054201},
  year={2016},
  publisher={APS}
}

@article{serbyn2013local,
  title={Local conservation laws and the structure of the many-body localized states},
  author={Serbyn, Maksym and Papi{\'c}, Zlatko and Abanin, Dmitry A},
  journal={Physical review letters},
  volume={111},
  number={12},
  pages={127201},
  year={2013},
  publisher={APS}
}

@article{chandran2015constructing,
  title={Constructing local integrals of motion in the many-body localized phase},
  author={Chandran, Anushya and Kim, Isaac H and Vidal, Guifre and Abanin, Dmitry A},
  journal={Physical Review B},
  volume={91},
  number={8},
  pages={085425},
  year={2015},
  publisher={APS}
}

@misc{OpenAINS2026,
  author       = {{OpenAI}},
  title        = {On the {N}avier--{S}tokes {M}illennium {P}rize {P}roblem},
  howpublished = {\url{https://openai.com/index/navier-stokes-solution/}},
  year         = {2026},
  note         = {Accessed 10 September 2026}
}

@article{broers2026exclusive,
  title={Exclusive-or encoded algebraic structure for efficient quantum dynamics},
  author={Broers, Lukas and Mathey, Ludwig},
  journal={New Journal of Physics},
  volume={28},
  number={4},
  pages={044508},
  year={2026},
  publisher={IOP Publishing}
}

@article{krotz2026paulib,
  title={PauLIB: A High-Performance Library for Processing Pauli Strings},
  author={Kr{\"o}tz, Florian},
  journal={arXiv preprint arXiv:2605.25974},
  year={2026}
}

@article{gopalakrishnan2019instability,
  title={Instability of many-body localized systems as a phase transition in a nonstandard thermodynamic limit},
  author={Gopalakrishnan, Sarang and Huse, David A},
  journal={Physical Review B},
  volume={99},
  number={13},
  pages={134305},
  year={2019},
  publisher={APS}
}

@article{gornyi2005interacting,
  author  = {Gornyi, I. V. and Mirlin, A. D. and Polyakov, D. G.},
  title   = {Interacting Electrons in Disordered Wires: Anderson Localization and Low-$T$ Transport},
  journal = {Physical Review Letters},
  volume  = {95},
  number  = {20},
  pages   = {206603},
  year    = {2005},
  doi     = {10.1103/PhysRevLett.95.206603}
}

@article{colmenarez2024ergodic,
  title={Ergodic inclusions in many-body localized systems},
  author={Colmenarez, Luis and Luitz, David J and De Roeck, Wojciech},
  journal={Physical Review B},
  volume={109},
  number={8},
  pages={L081117},
  year={2024},
  publisher={APS}
}

@article{sels2021dynamical,
  title={Dynamical obstruction to localization in a disordered spin chain},
  author={Sels, Dries and Polkovnikov, Anatoli},
  journal={Physical Review E},
  volume={104},
  number={5},
  pages={054105},
  year={2021},
  publisher={APS}
}

@article{abanin2021distinguishing,
  title={Distinguishing localization from chaos: Challenges in finite-size systems},
  author={Abanin, Dmitry A and Bardarson, Jens H and De Tomasi, Giuseppe and Gopalakrishnan, Sarang and Khemani, Vedika and Parameswaran, Siddharth A and Pollmann, Frank and Potter, Andrew C and Serbyn, Maksym and Vasseur, Romain},
  journal={Annals of Physics},
  volume={427},
  pages={168415},
  year={2021},
  publisher={Elsevier}
}

@article{de2013ergodicity,
  title={Ergodicity breaking in a model showing many-body localization},
  author={DeLuca, Andrea and Scardicchio, Antonello},
  journal={Europhysics Letters},
  volume={101},
  number={3},
  pages={37003},
  year={2013},
  publisher={EDP Sciences, IOP Publishing and Societ{\`a} Italiana di Fisica}
}

@article{abanin2019colloquium,
  title={Colloquium: Many-body localization, thermalization, and entanglement},
  author={Abanin, Dmitry A and Altman, Ehud and Bloch, Immanuel and Serbyn, Maksym},
  journal={Reviews of Modern Physics},
  volume={91},
  number={2},
  pages={021001},
  year={2019},
  publisher={APS}
}

@article{abou1973selfconsistent,
  title={A selfconsistent theory of localization},
  author={Abou-Chacra, R. and Anderson, P. W. and Thouless, D. J.},
  journal={Journal of Physics C: Solid State Physics},
  volume={6},
  number={10},
  pages={1734},
  year={1973}
}

\end{document}